\documentclass[12pt, preprint,twoside]{aastex}
\usepackage{geometry}               
\usepackage{graphicx}
\usepackage{amssymb}
\usepackage{epstopdf}
\usepackage{fullpage}
\usepackage{natbib}
\usepackage{subfigure}
\usepackage{epsfig}
\usepackage{multirow}
\usepackage[hang]{footmisc}
\setcitestyle{round,aysep={,},yysep={;}}
\title{The Effect of Host Star Spectral Energy Distribution and Ice-Albedo Feedback on the Climate of Extrasolar Planets}
\author{Aomawa L. Shields\altaffilmark{1,2,3},
Victoria S. Meadows\altaffilmark{2,3},
Cecilia M. Bitz\altaffilmark{4},\\ 
Raymond T. Pierrehumbert\altaffilmark{5},
Manoj M. Joshi\altaffilmark{6},
Tyler D. Robinson\altaffilmark{2,3}\\
}

\altaffiltext{1}{\bf{Corresponding author: Aomawa Shields, Department of Astronomy, University of Washington, Box 351580, Seattle, WA 98195-1580. aomawa@astro.washington.edu. \\Cell: 310-600-0751. Fax: 206-685-0403.}}
\altaffiltext{2}{Department of Astronomy and Astrobiology Program, University of Washington}
\altaffiltext{3}{NASA Astrobiology Institute}
\altaffiltext{4}{Department of Atmospheric Sciences, University of Washington}
\altaffiltext{5}{Department of Geological Sciences, University of Chicago}
\altaffiltext{6}{School of Environmental Sciences, University of East Anglia}

\begin{document}

\begin{abstract}
Planetary climate can be affected by the interaction of the host star spectral energy distribution with the wavelength-dependent reflectivity of ice and snow.  Here we explore this effect using a one dimensional (1-D), line-by-line, radiative-transfer model to calculate broadband planetary albedos as input to a seasonally varying, 1-D energy-balance climate model. A three-dimensional (3-D) general circulation model is also used to explore the atmosphere's response to changes in incoming stellar radiation, or instellation, and surface albedo. Using this hierarchy of models we simulate planets covered by ocean, land, and water ice of varying grain size, with incident radiation from stars of different spectral types. Terrestrial planets orbiting stars with higher near-UV radiation exhibit a stronger ice-albedo feedback. We find that ice extent is much greater on a planet orbiting an F-dwarf star than on a planet orbiting a G-dwarf star at an equivalent flux distance, and that ice-covered conditions occur on an F-dwarf planet with only a 2\% reduction in instellation relative to the present instellation on Earth, assuming fixed CO$_2$ (present atmospheric level on Earth). A similar planet orbiting the Sun at an equivalent flux distance requires an 8\% reduction in instellation, while a planet orbiting an M-dwarf star requires an additional 19\% reduction in instellation to become ice-covered, equivalent to 73\% of the modern solar constant. The reduction in instellation must be larger for planets orbiting cooler stars due in large part to the stronger absorption of longer-wavelength radiation by icy surfaces on these planets, in addition to stronger absorption by water vapor and CO$_2$ in their atmospheres, providing increased downwelling longwave radiation. Lowering the infrared and visible band surface ice and snow albedos for an M-dwarf planet increases the planet's climate stability to changes in instellation, and slows the descent into global ice coverage. The surface ice-albedo feedback effect becomes less important at the outer edge of the habitable zone, where atmospheric CO$_2$  can be expected to be high in order to maintain clement conditions for surface liquid water. We show that $\sim$3-10 bars of CO$_2$ will entirely mask the climatic effect of ice and snow, leaving the outer limits of the habitable zone unaffected by the spectral dependence of water ice and snow albedo. However, less CO$_2$ is needed to maintain open water for a planet orbiting an M-dwarf star, than would be the case for hotter main-sequence stars.\end{abstract}
\keywords{EXTRASOLAR PLANETS, M STARS, HABITABLE ZONE, SNOWBALL EARTH}
\maketitle
\pagebreak

\section{Introduction}
M-dwarf stars are believed to comprise $\sim$75\% of all main-sequence stars, therefore they offer the best chance to find orbiting habitable planets through sheer numbers alone. M-dwarf hosts also demonstrate clear advantages for planetary habitability. Because of their small masses compared to the Sun (0.08-0.5 M$_\odot$), M-dwarf stars burn nuclear fuel at a much slower rate, and are therefore extremely long-lived, offering lengthy timescales for planetary and biological evolution. Earth-sized planets around M-dwarf stars are easier to detect with existing techniques, since both stellar radial velocity and photometric transit depths are larger given the lower star-to-planet mass and size ratios \citep{Tarter2007}. Because of the lower luminosities of M-dwarfs, their habitable zones (defined as the locus of orbits around a star where an Earth-like planet can support liquid water on its surface, \citealp{Kasting1993}) will be much closer to the star for M-dwarfs (0.24-0.44 AU for an early-type M-dwarf, assuming an effective temperature of 3800 K and luminosity of 0.05 L$_\odot$, \citealp{Kopparapu2013}) than for G-dwarf stars like the Sun (0.99-1.70 AU, \citealp{Kopparapu2013}), increasing the geometric probability of observing a transit, which scales inversely with a planet's orbital distance from its host star \citep{Borucki1984}. 

A planet orbiting an M-dwarf star (henceforth called an M-dwarf planet) would experience a very different environment from that of a planet orbiting a larger, brighter star. In such close-in orbits, tidal effects are expected to be strong, and could lead to synchronous rotation, where the planet takes as long to make one rotation as it does to complete one orbit around its star \citep{Dole1964, Kasting1993, Joshi1997, Edson2011}. Previous work suggests that aqua planets cool with increasing rotation period, due to weakened low-latitude zonal winds that allow sea ice to expand across the planet \citep{Edson2011}. Initial concerns about M-dwarf planet atmospheres condensing out on the permanent night side were assuaged by climate simulations of synchronously-rotating M-dwarf planets, which suggest that 0.1 bar of a gas such as CO$_2$ is enough to ensure sufficient heat transport between the sun-lit and dark sides of the planet, preventing atmospheric freeze-out \citep{Joshi1997, Edson2011}. Edson \emph{et al.} (\citeyear{Edson2011}) also found that increased atmospheric latent and oceanic diffusive heat transport reduced temperature contrasts between the sunlit and dark sides of an aqua planet.  M-dwarf stars also exhibit high degrees of chromospheric activity, resulting in flares that emit significant amounts of high-energy radiation \citep{Hawley1991}, which could be detrimental to surface life. However, photochemical and radiative/convective models have yielded a maximum depletion in atmospheric ozone number density of 1\% due to UV radiation emitted by flares, indicating that planetary surface habitability may not be unequivocally threatened by stellar flare events. The addition into simulations of the proton flux that can be associated with stellar flares results in a reduction in ozone column depth by as much as 93\%. This is an upper limit, as not all flares are accompanied by the ejection of energetic ionizing particles, and the effect of flare-associated particle ejections on atmospheric ozone concentration could be ameliorated by a planet's magnetic field \citep{Segura2010}. 

M-dwarf stars emit a large fraction of their radiation at near-infrared (near-IR) wavelengths, which can strongly affect planetary climate. About 95\% of the radiation emitted by the M-dwarf star AD Leo is at wavelengths longer than 0.7 $\mu$m, compared with $\sim$53\% for the Sun. Given the dependence of ice and snow albedo on wavelength, the interaction between the spectral energy distribution (SED) of M-dwarf stars and the icy or snowy surfaces that may exist on their orbiting planets will also differ greatly from that of other host stars and their planets' ice surfaces. In the near-UV (0.3 $\mu$m $< \lambda <$ 0.4 $\mu$m) and visible (0.4 $\mu$m $\leqslant \lambda \leqslant$ 0.7 $\mu$m) range of the electromagnetic spectrum, both ice and snow albedo are high, while in the infrared (IR) and near-IR ($\lambda >$ 0.7 $\mu$m), the reflectance of ice and snow drops substantially, due to an increase in their absorption coefficients at longer wavelengths \citep{Dunkle1956}; a consequence in large part of molecular vibrations involving various combinations of the water molecule's three fundamental vibrational modes \citep{Farrell1967}. Because of the spectral dependence of ice albedo, the ice-albedo feedback mechanism is sensitive to the wavelength of light coming from the host star. On M-dwarf planets, a significant amount of radiation emitted by the host star is absorbed by atmospheric gases such as CO$_2$ and water vapor, which absorb strongly in the near-IR \citep{Kasting1993, Selsis2007}. However, a disproportionate amount of the longer-wavelength radiation that does reach the surface will be absorbed by, rather than reflected from, icy or snowy surfaces on these planets \citep{Joshi2012}. This will reduce the difference between ice and ocean surface albedo. Episodes of low-latitude glaciation,  termed ``Snowball Earth'' events \citep{Kirschvink1992} may be less likely to occur on M-dwarf planets as a result of the lower-albedo ice on their surfaces, as entrance and exit into such a snowball state has been shown to be sensitive to ocean-ice albedo contrast \citep{Pierrehumbert2011b}.

Initial work in this area has been of a purely analytical nature, using synthetic stellar and blackbody spectra, one surface water ice type and one snow type to show that broadband albedos are likely lower for M-dwarf planets \citep{Joshi2012}. Neither radiative transfer through an atmosphere, nor the effect of clouds, were included in these calculations. The effect of near-IR CO$_2$ and water vapor absorption was estimated using a constant scaling factor at wavelengths beyond 1.5 $\mu$m. Here, we provide a more comprehensive study of this phenomenon, using a hierarchical approach. We use a line-by-line, radiative-transfer model that is 1-D in altitude \citep{Meadows1996, Crisp1997}, an energy balance model (EBM) that is 1-D in latitude \citep{North:1979fk}, and a 3-D general circulation model (GCM) with both horizontal and vertical spatial resolution \citep{Gent2011}. We also use wavelength-dependent albedo spectra for five different surface ice and snow types across the full near-UV through near-IR range (\citealp{Grenfell1994, Warren2002}), with empirical stellar spectra for F-, G-, K-, and M-dwarf stars (\citealp{Segura2003,Segura2010,Chance2010}).

Historically, the inner and outer edges of the habitable zone (henceforth IHZ and OHZ) are often defined by the boundary conditions of the runaway greenhouse, and the planetary orbital distance beyond which an increase in atmospheric CO$_2$ concentration is no longer sufficient to maintain surface temperatures above the freezing point of water, termed the maximum CO$_2$ greenhouse \citep{Kasting1993, Underwood2003, Pierrehumbert2010, Kane2012, Kopparapu2013}. Joshi and Haberle (\citeyear{Joshi2012}) speculated that the OHZ, which they define as the orbital distance at which carbon dioxide condenses out onto the surface of a planet, may be as much as 30\% farther out for M-dwarf stars than originally calculated, given the spectral dependence of water ice and snow albedo. However, at the very outer limit of the habitable zone, atmospheric CO$_{2}$ content is expected to be high as a result of decreased silicate weathering \citep{Walker1981}. As the maximum greenhouse limit for CO$_{2}$ (more than $\sim$2 bars, \citealp{Pierrehumbert2010}) is approached, planetary albedo becomes dominated by the thicker CO$_{2}$ atmosphere, and can reduce climate sensitivity to surface ice albedo. The exact CO$_{2}$ concentration required to mask the climatic effect of the interaction between host star SED and surface ice and snow has not been constrained, and we determine it here. 

The wavelength dependence of ice and snow albedo has been derived analytically assuming diffuse solar irradiation \citep{Dunkle1956}. Subsequent model comparisons with field measurements have verified the spectral behavior of ice and snow albedo, and highlighted its dependence on directional irradiation and solar zenith angle, as well as grain size, with absorption and forward scattering of radiation increasing for larger grains \citep{Wiscombe1980}. Here we explore a large parameter space of possible ice grain sizes and densities to examine the effect of these surface types on planetary climate.  

In subsequent sections, we calculate the effect of stellar SED on planetary climate using a three-tiered approach. First, we use a 1-D radiative transfer model, the Spectral Mapping Atmospheric Radiative Transfer model (SMART), to investigate the interaction between the spectrum of a host star and that of multiple ice types, land, and ocean on a planet orbiting the star at the distance at which it receives an equivalent amount of flux to that which Earth receives from the Sun (hereafter ``equivalent flux distance"). Second, we use a 1-D EBM with our SMART-derived broadband planetary albedos as input, to examine the effect of the interaction between the host star's incident radiation (hereafter ``instellation") and the latitudinal distribution of surface temperature and ice extent, as a function of the energy balance of the planet. Lastly, we employ a more sophisticated, 3-D GCM to explore the response of the atmosphere and its interaction with the ocean, given the SED of the host star. We also use the 3-D GCM to confirm general trends implied by our EBM simulations, and to constrain the effect of the spectral dependence of surface water ice and snow albedo on planetary climate. For all model simulations, we have assumed a rotation period similar to Earth's, in order to isolate the effects of SED on planetary climate. As M-dwarf planets are eventually likely to become synchronously rotating \citep{Dole1964, Kasting1993, Joshi1997, Edson2011}, we look forward to exploring this effect in future work. 

\section{Models}
Here we describe our method of using a 1-D radiative transfer model to calculate broadband planetary albedos assuming different incident stellar spectra, a wide range of ice and snow surface types, and land and ocean as input to a seasonally-varying, 1-D EBM. With the EBM we explore the latitudinal extent of the ice line (defined here as the farthest latitude towards the equator where ice is present on a planet's surface), and thus what fraction of the planet is ice-covered, as a function of incident stellar and surface albedo spectrum. Using our radiative transfer model, with spectrally-resolved absorption coefficients for CO$_2$, we determine the atmospheric CO$_2$ concentration necessary to mask surface ice-albedo feedback effects entirely. The manner in which radiative transfer through atmospheric gases is treated can affect our EBM results, and Earth's temperature and ice cover extent have been shown to be greatly influenced by variations in solar radiation \citep{Budyko1969}. To address these effects, we run a 3-D atmospheric GCM coupled to a motionless 50m-depth ``mixed-layer" ocean with a thermodynamic sea ice model. There is no land, and hence we refer to it as an aqua planet. We vary the SED of the instellation in order to investigate the climate of a hypothetical planet of a host M-dwarf star, and compare it to the climate of an aqua planet orbiting the Sun, a G-dwarf star, as well as an F-dwarf star, which emits a greater fraction of visible and near-UV radiation. We also describe how changes to the surface albedo parameterization in the GCM affect the climate of an M-dwarf planet. We discuss implications of the results of our climate model simulations for M-dwarf habitability, given the effect of host star SED on climate sensitivity. 

\subsection{SMART}
SMART is a 1-D, multistream, multilevel, line-by-line, multiple-scattering radiative transfer model developed by David Crisp \citep{Meadows1996, Crisp1997}. It is used to generate high-resolution synthetic spectra of planetary atmospheres and surfaces. Recent upgrades and modifications have been made to SMART, including the implementation of Jacobians to support climate modeling. The current version of SMART has been used most recently by Robinson \emph{et al.} (\citeyear{Robinson2011}) to model the Earth. We have used SMART with updated gas absorption line profiles as described in Section 2.4.3. The model uses data on the stellar spectrum, planetary atmospheric composition and temperature, and surface and cloud properties as inputs, and computes wavelength-dependent radiative intensity and flux profiles, for both reflected and emitted light. From the computed specific fluxes, we calculate bolometric upwelling and downwelling stellar flux profiles, which we then use to generate the total broadband planetary albedo given the input stellar and surface albedo spectrum.

\subsection{Energy Balance Model}
Energy balance climate models (EBMs) simulate the climate from an equation of energy balance of the planet. We used a seasonally varying, 1-D EBM adapted from North and Coakley (\citeyear{North:1979fk}).  The energy balance at each latitude is the sum of absorbed shortwave radiation, outgoing longwave radiation, and the convergence of horizontal heat transport equated with the heat tendency of the vertical column. 

We began by coding up the North and Coakley model (\citeyear{North:1979fk}) ourselves, and verified that it reproduced the results of their paper. We then modified it as follows: Heat transport is computed from diffusion of temperature with a latitudinally-varying diffusion coefficient, which enhances tropical relative to subtropical heat transport, thus smoothing temperature variations in the tropics and leading to consistency with thermal wind observations \citep{Lindzen1977}.  We used idealized land, and prescribed time-independent orbital parameters. The instellation is spectrally integrated, and varies with latitude and time of year. We also varied the albedo of land, ocean, and below-freezing surfaces, depending on the SED of the instellation. The planetary albedos calculated using SMART with different stellar spectra, an Earth-like atmosphere, and surface albedo files for land, ocean, and ice of varying grain size are input as broadband albedos over land and ocean in the EBM. This allowed the EBM to incorporate to first order the effect of varying SED on the energy balance of the planet.  This EBM configuration assumes a rapidly rotating planet like the Earth. More work therefore remains to be done to examine alternate spin states such as synchronous rotation, which is expected to prevail in the habitable zones of M-dwarf stars.

There is no explicit ice model; instead the ocean temperature is allowed to be below freezing, which then causes the albedo to increase. We acknowledge that neglecting the energy- and time-dependent dynamics of melting ice will affect the seasonal cycle, therefore we use the EBM primarily to determine the annual mean temperature behavior, and defer to the results of our GCM simulations regarding the behavior of the ice margin. 

The dependence of outgoing longwave radiation on temperature is linearly parameterized based on the general effect of CO$_{2}$ and water vapor on atmospheric radiative transfer as a function of planetary surface temperature. Each latitude has a land and ocean fraction specified to be uniform at 0.01 and 0.99, respectively, in order to approximate an aqua planet. We evaluated how the ice line latitude changed as a function of modern stellar flux, given the broadband planetary albedos over land, ocean, and ice-covered areas that were calculated with SMART using different stellar spectra as input. 

\subsection{General Circulation Model}
We used version 4.0 of the Community Climate System Model (CCSM), a fully-coupled, global climate model developed to simulate past, present, and future climate states on the Earth \citep{Gent2011}. We ran CCSM4 with a 50-meter deep, slab ocean (see e.g., \citealp{Bitz2012}). The slab ocean model normally has ocean heat flux convergence (often called a q-flux) input. We set this flux to zero, as done in experiments by Poulsen \emph{et al.} (\citeyear{Poulsen2001}), though we acknowledge that this allows for a snowball transition at a higher instellation than if ocean dynamic heat flux, which tends to buffer against a snowball state, were incorporated \citep{Poulsen2001, Pierrehumbert2011b}. The ocean is treated as static, but fully mixed with depth. Simulations that include a fully-dynamic ocean, though more realistic, can take thousands of model years to equilibrate for a given radiative forcing, and as such are too computationally expensive to allow for the exploration of a wide parameter space as we do here. 

The sea ice component is the Los Alamos sea ice model CICE version 4 \citep{Hunke2008}, except we have made the ice thermodynamic only (no sea-ice dynamics, which may affect the snowball transition, \citealp{Abbot2011}), and have reverted back to the sea-ice albedo parameterization from CCSM3. The latter change was made because the newer Delta-Eddington, multiple scattering albedo parameterization is much more difficult to control. In the CCSM3 albedo parameterization, the surface albedo is divided into two bands, visible ($\lambda \leqslant$ 0.7 $\mu$m) and near-IR ($\lambda >$ 0.7 $\mu$m). Default near-IR and visible wavelength band albedos are 0.3 and 0.67 for cold bare ice and 0.68 and 0.8 for cold dry snow. The albedos decrease modestly as the surface approaches the melting temperature to mimic the effect of melt ponds. The CCSM3 cold bare ice albedo is approximately halfway between the two end-members for blue marine ice and snow in Figure 1, while the CCSM3 cold dry snow albedo is between the brightest end-member, and 3/4 of the way towards the brightest end-member in Figure 1. We use the default near-IR and visible band albedos for all simulations except the surface sensitivity tests described in Section 4.2.4. However, near-IR and visible band ice albedos calculated using our ice and snow spectra and weighted by the M-dwarf SED are higher than the default values by 3.3\% and 1.5\%, respectively, higher by 21\% for the visible band snow albedo, and lower by 23\% for the near-IR snow albedo. An implicit error therefore must exist within this range as a result of using the default GCM two-band model, and may affect the exact values of temperature and ice extent we have calculated for different SEDs. But as the spectral dependence of ice and snow albedo is still captured in the two-band model, the general trends shown here are expected to be robust. 

The Community Atmosphere Model version 4 (CAM4.0) divides the incident shortwave (stellar) radiation into twelve wavelength bands, given in Table 1. The percentage of the total flux from the star is specified in these wavelength bands, with a significantly higher percentage of total stellar flux at longer wavelengths (94.6\% for M-dwarf star AD Leo, vs. 52.8\% for the Sun, and 40.7\% for F-dwarf star HD128167 in waveband 10). As AD Leo emits non-negligible amounts of radiation at wavelengths shortward of 0.2 $\mu$m and longward of 5.0 $\mu$m, the percentages of flux in these regions were folded into the shortest (Band 1) and longest (Band 10) wavebands, respectively, to include the full stellar spectrum. 

We have assumed an Earth-like atmospheric composition for CO$_{2}$, CH$_{4}$, H$_{2}$O, and O$_{2}$. The ozone profile has been set to zero, as we are concerned primarily with surface temperature as a measure of planetary habitability, and have found the presence of ozone to have a negligible effect on the surface temperature of M- and G-dwarf planets. F-dwarf planets may exhibit colder surface temperatures and greater ice extent with ozone in their atmospheres than without, possibly as a result of greater stratospheric heating in F-dwarf planet atmospheres  \citep{Segura2003} preventing more shortwave radiation from reaching the surface. 

We ran the model on four intel Xeon hex core chips at 2.27 GHz, giving $\sim$11 simulation years per wallclock day. For each GCM simulation, the model was run for 37 years to equilibrate to the modern Earth climate at present solar instellation. We then ran our simulations for 40 years after that, with varying percentages of instellation with the SED from either the Sun (a G-dwarf star), an M-dwarf star, or an F-dwarf star. 

\subsection{Model Inputs}

\subsubsection{Stellar Spectra}
Earlier GCM simulations explored the albedo effects of surface water ice on the climate of M-dwarf planets using a blackbody distribution to approximate incoming stellar radiation \citep{Pierrehumbert2011a}. Here we use composite SEDs derived from observations and models of main-sequence (core hydrogen-burning) stars of different spectral types to calculate broadband planetary albedos for a planet covered in specific surface types. We used F2V Star HD128167, K2V Star HD22049 \citep{Segura2003}, M3V star AD Leo\footnotemark{} (\citealp{Reid1995}, \citealp{Segura2005}), and the solar spectrum obtained from Chance and Kurucz (\citeyear{Chance2010}). All SEDs are shown normalized by their peak flux in Figure 1. For our SMART runs, all stellar spectra were adjusted so as to integrate to top-of-atmosphere solar constant values of between 1360 and $\sim$1380 W/m$^2$. For our GCM runs, the value of the solar constant was set to 1360 W/m$^2$ for all SEDs. This is equivalent to placing our model planets at the Earth's equivalent distance from their host stars. The zenith angle varies according to the diurnal cycle in the GCM, and in our SMART runs it is fixed at an average illumination value of 60$^{\circ}$.

\footnotetext[1]{http://vpl.astro.washington.edu/spectra/stellar/mstar.htm}

\subsubsection{Surface Albedos}
Previous EBMs have employed a broadband frozen surface albedo of 0.6 for M-dwarf star illumination \citep{Pierrehumbert2011a}. Here, we present a more comprehensive look at the effect of incoming stellar radiation on the broadband frozen surface albedo, using high-resolution, wavelength-dependent albedo spectra for multiple water ice types. We use spectra of albedo as a function of wavelength for fine-grained snow \citep{Grenfell1994} and for blue marine ice \citep{Warren2002}. Bubbles in glacier ice result from air trapped in snow as it is compressed under the weight of overlying snow. The blue marine ice was not glacier ice; it resulted instead from freezing of liquid water. It contained few bubbles but numerous cracks, which were responsible for the albedo. The cracks are caused by thermal stresses, so they would be present (and contributing to the albedo) even on a planet with no atmosphere \citetext{Stephen Warren, priv.\ comm.}. Since these two ice types constitute end-members in terms of ice grain size and albedo at Earth-like surface temperatures and pressures, and ice may exist between these two states, wavelength-dependent albedo spectra have been modeled at 25\%, 50\%, and 75\% mixtures of blue ice and snow (Fig. 1). Albedos for $\lambda < $ 0.31 $\mu$m were decreased linearly from their values at 0.31 $\mu$m down to 0.05 at 0.15 $\mu$m for all ice and snow types.

As clouds also exhibit a wavelength-dependent reflectivity, in some cases reaching higher albedos than snow surfaces at longer wavelengths, we plot the contribution of two major cloud types to the Earth's overall reflectivity (Fig. 2), and include them in our model validation in Section 3. Clouds, particularly low stratocumulus clouds, have the largest contribution to the Earth's reflectivity at the longest wavelengths of the shortwave spectrum (i.e. 1.5 - 2.5 $\mu$m), where the reflectivity of snow and high cirrus cloud drop significantly. We assume an underlying snow surface in Figure 2 in order to highlight the contrast in the wavelength-dependent reflectivity between snow and clouds.

We use an ocean albedo spectrum obtained from Brandt \emph{et al.} (\citeyear{Brandt2005}). For wavelengths between 0.15 and 0.31 $\mu$m, albedos were calculated using the Fresnel equations, with indices of refraction of water from Segelstein (\citeyear{Segelstein:Thesis:1981}). Following Robinson \emph{et al.} (\citeyear{Robinson2011}), we obtained the spectrum for the clay mineral kaolinite (mixed with trace amounts of smectite and illite) from the United States Geological Survey (USGS) spectral library\footnotemark{} for use as the bare land surface in our EBM simulations.

\footnotetext[2]{http://speclab.cr.usgs.gov/spectral-lib.html}

\subsubsection{Absorption Coefficients}
We generated absorption cross sections for relevant gases using the HITRAN 2008 line list database\footnotemark{} \citep{Rothman2009}. We then simulated line profiles using a line-by-line absorption coefficient model (LBLABC, \citealp{Meadows1996}).

\footnotetext[3]{http://www.cfa.harvard.edu/hitran/}

\section{Model Validation}
We validated the method of using SMART in combination with an EBM by reproducing the Earth's current ice line latitude and global mean surface temperature at its present obliquity, using outgoing infrared flux and diffusive heat transport parameterizations as described by North and Coakley (\citeyear{North:1979fk}). Our maximum and minimum top-of-atmosphere upward fluxes match observations from the Earth Radiation Budget Experiment (ERBE, \citealp{Barkstrom1982}) to within 6\%. We used a temperature-pressure profile for the atmosphere derived using the Intercomparison of Radiation Codes in Climate Models program (ICRCCM). The profile for ICRCCM mid-latitude summer sounding includes atmospheric pressures, temperatures, and mass mixing ratios for seven absorbing gases, and is resolved into 64 vertical layers throughout the atmosphere. We included absorption by H$_{2}$O, CO$_{2}$, O$_{3}$, N$_{2}$O, CO, CH$_{4}$, and O$_{2}$, and Rayleigh scattering. Broadband planetary albedos for the surface types described in Section 2 were generated assuming both clear and cloudy sky conditions. Approximately 64\% of the Earth is covered in clouds \citep{Hahn2002}. Two cloud layers were used - that of low stratocumulus clouds and high cirrus clouds. For the purposes of our model, we assume 36\% clear sky conditions, 40\% low stratocumulus clouds, and 24\% high cirrus clouds.  Average broadband albedos over ocean, land, and regions where the surface temperature falls below -2$^\circ$C were calculated as described in Zsom \emph{et al.} (\citeyear{Zsom2012}), using these cloud fractions as well as the combined cloud fraction assuming random overlap between the two cloud layers \citep{Oreopoulos2003}. An optical depth of 10 was used for both cloud types. A modern land/ocean geographical configuration was used in the EBM. Areas over ocean and land surfaces were assigned broadband albedos of 0.32 and 0.41, respectively.  Regions where the surface temperature fell below -2$^\circ$C were assigned a broadband albedo of  0.46, which corresponds to the 25\% mixture of blue marine ice and fine-grained snow in Figure 1. The EBM ice line latitude is 54.4$^\circ$N, which is within six degrees of the Earth's northern hemisphere ice line latitude as defined as the southernmost tip of Greenland by Kukla (\citeyear{Kukla1979}) and Mernild \emph{et al.} (\citeyear{Mernild2010}).  The EBM global mean surface temperature (including zonal mean land and ocean temperatures) is 12$^\circ$C, and is within three degrees of the global mean surface temperature of the Earth as defined as 15$^\circ$C by Hartmann (\citeyear{Hartmann1994}). The EBM ice line latitude and global mean surface temperature (Fig. 3) were calculated using only the surface types in Figure 1. 

We have assumed a water vapor concentration equal to a mid-latitude average of 1\% of the atmosphere at 1-bar surface pressure to calculate broadband planetary albedos used in our model validation. The amount of water vapor decreases rapidly with increasing latitude, and can be as much as 10 times lower at the poles \citep{Hartmann1994}. To test the sensitivity of our calculated albedos to water vapor, we ran SMART with 0.1\% water vapor, and found our broadband albedos over land, ocean, and below-freezing surface types to increase by no more than 10\% for ocean and ice surfaces, and 13\% for our land spectrum.

\section{Results} 

\subsection{SMART + EBM}
We used SMART to comprehensively address the radiative transfer that takes place between a planet's ice, ocean, and land surfaces and the incident stellar radiation. Derived broadband planetary albedos used as input to the EBM are therefore unique to the planet-star system, rather than broadly generalized. Here we first explore the contributions to broadband planetary albedo for different surface types, atmospheric pressures, and stellar SEDs. We then explore climate-sensitivity to changes in instellation, surface type, and ice grain particle size for Earth-like planets orbiting stars of different spectral type. 

\subsubsection{Contributions to Broadband planetary albedo}

Figure 4 shows broadband planetary albedos that are output from SMART, given various surface types (where broadband, or ``Bond'' albedo is defined as the ratio of the total amount of flux reflected by the planet, divided by the total flux incident on the planet, integrated over the entire wavelength spectrum).

We are primarily interested in quantifying the effects of surface albedo on planetary climate, and used the aqua planet EBM as a testbed for identifying general trends. Broadband planetary albedos decrease monotonically with stellar luminosity, even though all stellar spectra have been adjusted to provide the same integrated flux as that received by the Earth around the Sun. This can be explained by the larger fraction of spectral energy emitted at $\lambda >$ 1.0 $\mu$m for stars with lower luminosities. These broadband planetary albedos incorporate the wavelength-dependent reflectivity of the underlying surface. At these longer wavelengths, ice and snow absorb strongly (Fig. 1), resulting in a lower average surface albedo contribution to the overall planetary albedo. 

Since Rayleigh scattering is stronger at shorter wavelengths due to the 1/$\lambda^4$ cross-section dependence, it is expected to be more pronounced on an F-dwarf planet than on other host star planets, and can result in higher broadband planetary albedos (\citealp{Kasting1993, Pierrehumbert2011a, Kopparapu2013}). To verify the trend stated above, we reduced the atmospheric pressure in additional SMART runs with an F-dwarf spectrum incident on an ocean surface, and excluded atmospheric gases. We confirm that the Rayleigh scattering tail present at $\lambda <$ 0.7 $\mu$m for atmospheres with 1 bar surface pressure is virtually absent for thin atmospheres of less than 0.1 mb (Fig. 5). The resulting broadband albedos in the ``no Rayleigh scattering'' case are plotted in Figure 4 (middle column). Even with Rayleigh scattering removed,  Figure 4 reveals higher broadband planetary albedos with ice and snow surfaces for the F-dwarf host star, and the lowest broadband albedos for the M-dwarf host star. Also plotted are broadband planetary albedos calculated with Rayleigh scattering only (no atmospheric gas absorption, left column), and with both Rayleigh scattering, atmospheric gases, and clouds included (right column). These values for each stellar type are listed in Table 2.

\subsubsection{Climate sensitivity for Earth-like planets }

Broadband planetary albedos calculated from SMART (with an Earth-like atmospheric composition and distribution of clouds) were used as input to the EBM. Below-freezing surfaces encountered during the EBM runs (where the temperature is less than -2 degrees Celsius) were given broadband planetary albedos for ice and snow of varying grain size calculated from SMART at 1 bar surface pressure. EBM simulations were run with an initial warm start, with an approximate Earth-like zonal mean temperature distribution. The global mean surface temperature and mean latitude of the ice line in the northern hemisphere were calculated at present Earth obliquity (23.5$^\circ$) as a function of percent of modern solar constant, with 100\% of the modern solar constant being the present amount of solar instellation, or flux density, on the Earth. The results are plotted in Figure 6, and can be assumed to be similar for the southern hemisphere, given an assumed eccentricity of zero. The slope of each line is a measurement of the climate sensitivity of the planet to changes in instellation. The smaller the grain size of the ice (and therefore the higher the albedo), the farther towards the equator the ice advances. For stars with higher visible and near-UV output, ice-covered planetary states occur with smaller decreases in instellation. Cooler, redder stars allow planets to remain free of global ice cover with larger decreases in instellation. It must be noted that an alternate branch of these results exists with significant ice cover if an initial cold start is assumed, with zonal mean temperatures characteristic of a snowball state. We have not explored those states here.

\subsection{GCM simulations}

1-D EBMs offer insight into the latitudinally-resolved distribution of temperature and ice across a planet. However, they heavily parameterize atmospheric radiative transfer and eddy heat transport. Furthermore, ours only captures the albedo effects of sea ice (not the insulating effect). To test the robustness of the results from our EBM simulations, and to provide further physical insight into the contribution of atmospheric dynamics to planetary climate, we employed a coupled atmospheric GCM with more complete sea ice physics. Coupled GCMs are crucial to understanding the full planetary response to changes in radiative and surface forcing. This is demonstrated in Figure 7, where the top-of-atmosphere absorbed shortwave radiation minus outgoing longwave radiation as a function of latitude in an EBM is compared with that in a GCM for a G-dwarf planet receiving 100\% of the modern solar constant. When averaged over a few years or more, the net incoming heat flux must be equal to the divergence of heat from each grid cell in order to yield a net surface flux of zero locally and globally in our slab ocean model. The EBM shows a large jump in the net incoming heat flux near the poles, due to the abrupt change in albedo for ice-covered areas in the EBM, and the lack of parameterized clouds (beyond our SMART treatment). This large jump causes the ice line to be unstable and collapse to the equator at significantly higher latitudes in the EBM compared to the GCM. A smoother transition in net incoming heat flux as a function of latitude is visible in the GCM, due to the presence of full-scale atmospheric dynamics, including clouds. The ice line latitude is stable 17-24 degrees farther towards the equator for all simulated aqua planets (Fig. 8) than those calculated in the EBM (Fig. 6), yielding gentler transitions in latitudinal ice extent and global mean surface temperature as a function of percent of modern solar constant in the GCM than is seen in the EBM. As the slopes of the lines in Figures 6 and 8 are a measure of the climate sensitivity of the planets, the GCM results indicate a lower climate sensitivity to changes in instellation than would be inferred from the EBM results alone.  

We begin our analysis of GCM simulations by characterizing how the global climate depends on the stellar constant for an aqua planet orbiting the Sun or an M-or F-dwarf star. Then we describe the climates in greater detail for an aqua planet around the Sun and an M-dwarf star, where we consider runs with similar temperate climates but vastly different SEDs, followed by stellar constants that give similar snowball climates. We explore the role played by atmospheric dynamics in generating similar ice-covered states, given the different stellar SEDs. Finally, we investigate the sensitivity of planetary climate to surface albedo alone.

\subsubsection{Climate sensitivity: G-dwarf vs. M-dwarf and F-dwarf planets}

Figure 8 shows the latitudinal ice extent and annual global mean surface temperature in the northern hemisphere as a function of percent of modern solar constant for a G-dwarf aqua planet with an incident spectrum of the Sun, an M-dwarf aqua planet with the incident spectrum measured for AD Leo, and an F-dwarf aqua planet with the incident spectrum measured for HD128167. As we have assumed an eccentricity of zero in our GCM runs, an obliquity of 23$^{\circ}$, and no land surface, the temperature and ice behavior are symmetric in the time-mean about the equator over an annual cycle. A comparison of 3-D simulations of the climate of a planet receiving 90\% of the modern solar constant (1224 W/m$^2$) from an M-dwarf with one receiving 90\% flux from the Sun yields significant differences in the resulting climate states. The planet receiving the Sun's SED is completely ice-covered, with a global mean surface temperature of 215 K, while the planet receiving the M-dwarf star's SED is 72 degrees hotter (at $\sim$288 K, see Fig. 8). This is largely due to the increased absorption of the predominantly near-IR radiation emitted by M-dwarf stars, by atmospheric CO$_2$ and water vapor. This results in a much larger greenhouse effect that keeps the M-dwarf planet's surface warmer than the G-dwarf planet \citep{Kasting1993, Selsis2007}. The M-dwarf planet's global mean surface temperature and ice line latitude (58.7$^{\circ}$) are in better agreement with the global mean surface temperature and latitudinal ice extent on a G-dwarf planet receiving 100\% of the modern solar constant ($\sim$1360 W/m$^2$), which are 287 K and 58.3$^{\circ}$, respectively. One-dimensional radiative-convective calculations by Kasting \emph{et al.} (\citeyear{Kasting1993}) yielded comparable stratospheric water vapor mixing ratios and planetary surface temperatures between G- and M-dwarf planets when the M-dwarf planet received 10\% less flux from its star compared to the G-dwarf planet, though the emission from the stars in their simulations was approximated by blackbody radiation \citep{Kasting1993}. 

With the CO$_{2}$ held fixed at 400 ppm, we started with simulations that would result in similar surface climate conditions on F-, G-, and M-dwarf aqua planets. On an F-dwarf planet, this required that the instellation be increased to 105\% of the modern solar constant in order to yield a comparable global mean surface temperature and ice line latitude (289 K and 59.8$^{\circ}$, respectively) to the G-dwarf planet receiving 100\% of the modern solar constant (287 K and 58.3$^{\circ}$). On an M-dwarf planet, an instellation equal to 90\% of the modern solar constant is all that is required to yield comparable a global mean surface temperature and ice line latitude (288 K and 58.6$^{\circ}$) to the G-dwarf planet receiving 100\% of the modern solar constant. 

On an F-dwarf planet receiving 100\% of the modern solar constant, the ice line latitude is $\sim$15$^{\circ}$. This is 43$^{\circ}$ farther towards the equator than the ice line latitude on a G-dwarf receiving equivalent instellation. A snowball state (where the ice covers the entire hemisphere, with a mean ice line latitude of 0$^{\circ}$) then occurs on the F-dwarf planet with a 2\% further reduction in instellation, to 98\% of the modern solar constant (Fig. 8). An 8\% reduction in instellation from the Sun is required to plunge the G-dwarf planet into global ice cover, at 92\% of the modern solar constant.  Ice on our M-dwarf planet does not extend all the way to the equator until the instellation has been reduced by an additional 19\% compared to the G-dwarf planet, equivalent to 73\% of the modern solar constant ($\sim$979 W/m$^2$). In Figure 8, the slope of the curve with decreasing stellar flux for the M-dwarf planet is much shallower than those for the G- and F-dwarf planets, indicating a much lower change in surface temperature and ice extent for a given change in stellar flux on the M-dwarf planet.  

As stated earlier, in our GCM simulations the ice line extends farther before unstable collapse to the equator than the critical latitude of 25-30$^{\circ}$ often required by diffusive EBMs. EBMs that transport less heat across the ice line, as well as fully-coupled atmosphere-ocean GCMs, have been found to permit stable ice lines at lower latitudes (\citealp{Poulsen2001, Pollard2005}). Cloud cover has also been shown to serve as a negative feedback on the latitudinal extent of tropical sea ice, as it contributes little additional reflectivity when located over high-albedo surfaces, but still adds to the greenhouse effect \citep{Poulsen2001}. Increased cloud cover may therefore contribute to stabilizing the ice edge in GCMs. And stable climate states with extremely low-latitude (but not equatorial) ice coverage on planets orbiting the Sun have been generated using climate models, as a result of large albedo differences between bare sea ice and snow-covered ice \citep{Abbot2011}. The fact that the ice line extends farthest towards the equator without collapsing to a snowball state on the M-dwarf planet than on the G- or F-dwarf planets is likely a consequence of the lower-albedo ice on the M-dwarf planet's surface. Indeed, the average albedo of the (mostly ice-covered) M-dwarf planet receiving 74\% of the modern solar constant (the lowest amount of instellation received by the planet in our simulations without becoming completely ice-covered) is 0.51, while the G- and F-dwarf planets with slightly less ice coverage (receiving 93\% and 99\% of the modern solar constant respectively), have mean albedos of 0.55 and 0.58. 

\subsubsection{90\% M-dwarf vs. 100\% G-dwarf: The climatic effect of stellar host SED}
The global mean surface temperature of the M-dwarf planet receiving 90\% of the modern solar constant is quite similar to a G-dwarf planet receiving the full modern solar constant (100\%), differing by less than one degree Kelvin. The latitudinal surface temperature profile is also strikingly similar (Fig. 9), although the M-dwarf planet exhibits warmer temperatures at the poles. In order to understand how the M-dwarf planet is compensating for the decreased instellation, we first take a closer look at the atmospheres of the two planets. 

While the surface temperature profiles of both planets appear relatively similar at sub-polar latitudes, there is about $\sim$10\% less precipitation on the M-dwarf planet everywhere except at the poles (Fig. 10). Precipitation is lower on the M-dwarf planet because there is less shortwave radiation reaching the surface to drive evaporation. 

The shift towards longer wavelengths in the M-dwarf star SED gives rise to a greater amount of shortwave heating due to absorption in the M-dwarf planet's atmosphere than on the G-dwarf planet, particularly in the tropics (Fig. 11a-c). The plot of zonal mean vertical temperature follows a similar pattern, with higher temperatures throughout most of the atmospheric column in the tropics and at the poles, indicating a smaller lapse rate over most of the troposphere on the M-dwarf planet than on the G-dwarf planet (Fig. 11d-f). The weaker Hadley circulation on the M-dwarf planet (Fig. 12) contributes to the greater tropospheric temperatures. However, near the surface, the difference in atmospheric temperature between the two planets is minimal. The stability of the atmospheres is therefore similar near the surface, but the M-dwarf planet appears more stable over the free troposphere, resulting in less cloudiness at most heights (Fig. 13), less precipitation, and more shortwave radiation absorbed by the planet. The more absorbing ice and snow on the M-dwarf planet (Fig. 14) allows the polar regions of the M-dwarf planet to be slightly warmer than the G-dwarf planet, with a similar ice line latitude despite the greatly reduced instellation.

\subsubsection{73\% M-dwarf vs. 92\% G-dwarf: A comparison of Snowball planets}
GCM simulations of an aqua planet receiving 73\% of the modern solar constant from an M-dwarf star and an aqua planet receiving 92\% of the modern solar constant from the Sun result in similar surface conditions, with ice extending all the way to the equator on both planets. A comparison of zonal mean vertical temperature on the two snowball planets shows higher temperatures in the upper troposphere of the M-dwarf planet (likely a consequence of greater shortwave heating due to absorption in the M-dwarf planet's atmosphere, as described in the previous section), but minimal difference in atmospheric temperature between the two planets near the surface (Fig. 15). Surface shortwave downwelling flux is the amount of flux from the host star that passes through the atmosphere and reaches the surface of the planet. Since the G-dwarf planet receives 19\% more radiation from its star than the M-dwarf planet, much more shortwave radiation reaches the surface of the G-dwarf planet (Figure 16a). However, the G-dwarf planet's surface is more reflective, due to the greater percentage of visible radiation (as discussed in Section 1) incident on the G-dwarf planet (Figure 16b). As a result the G-dwarf planet cannot absorb enough shortwave radiation to avoid full ice cover. Higher absorption of the shortwave spectrum incident on the surface of the M-dwarf planet compensates for the reduced instellation, resulting in similar amounts of shortwave radiation absorbed on the surfaces of both planets (Figure 16c), and similar global mean surface temperatures of $\sim$218 K (Figure 16d). Indeed, of the total shortwave flux that reaches the surface of the M-dwarf planet, $\sim$44\% is absorbed by the surface at the equator and $\sim$30\% is absorbed at the poles, compared with $\sim$35\% absorbed at the equator and $\sim$25\% at the poles on the G-dwarf planet. The three peaks in surface absorbed shortwave in the tropics and subtropics on the G-dwarf planet in Figure 16c result from local minima in the surface albedo (Fig. 16b) where evaporation exceeds precipitation and hence snow depths are locally smaller. 
 
 \subsubsection{Testing climate sensitivity to surface albedo}
Since the atmosphere of the M-dwarf planet absorbs such a large amount of the shortwave radiation, lowering the overall planetary albedo, it was important to verify whether the spectral dependence of surface ice albedo had any discernible effect on a planet's climate at Earth-like atmospheric CO$_2$ levels. By lowering the IR and visible band cold ice and snow albedos for an M-dwarf planet to 0.2, thereby significantly reducing the difference between ice and ocean on the planet's surface, we were able to examine the effect on climate sensitivity of surface ice and snow albedo interacting with the host star SED. Ice and snow albedos were lowered on M-dwarf planets receiving 75\% and 85\% of the modern solar constant, and the results were compared with model runs with the default albedo parameterization (see Section 2.3) for planets receiving equivalent instellation. The results are plotted in Figure 17. 

The result for both M-dwarf planets is a warmer climate and less ice than with the default albedo parameterization. The difference in climates is larger between the M-dwarf planets receiving 75\% instellation than between the M-dwarf planets receiving 85\% instellation, as indicated by the black vertical lines in Figure 17. This is because there is a higher surface coverage of ice on the M-dwarf planet receiving lower instellation, resulting in a larger impact on the planet's climate when the ice and snow albedos are lowered, including larger increases in temperature, cloudiness and precipitation. As a result, the M-dwarf planet's slope of the temperature and ice extent as a function of instellation is much shallower than that of the M-dwarf planet runs with the default albedo parameterization. This shallower slope implies a smaller change in surface temperature and ice extent for a given change in instellation, indicating that a planet's climate is sensitive to surface ice and snow albedo, even with an Earth-like atmospheric concentration of greenhouse gases. 

\subsection{High-CO$_2$ atmosphere: The outer edge of the habitable zone }
Atmospheric CO$_2$ concentrations are expected to increase farther out in the habitable zone, where decreased silicate weathering would lead to CO$_2$ building up in the atmosphere \citep{Walker1981}. CO$_2$ content at the OHZ would therefore be high, and could reach levels of over 2 bars before reaching the maximum CO$_2$ greenhouse limit \citep{Pierrehumbert2010}. Given that CO$_2$ has a number of absorption bands in the near-IR, where M-dwarf stars emit strongly and snow albedo drops significantly, we ran sensitivity tests to quantify the effect of CO$_2$ on the suppression of the cooling effect of ice-albedo feedback for M-dwarf planets. Runs were completed using SMART and included a case for a pure nitrogen atmosphere with little to no near-IR gas absorption, and cases for atmospheres consisting of 10\% CO$_2$ and 90\% N$_2$, 50\%/50\% CO$_2$/N$_2$, and pure CO$_2$ atmospheres at 2-, 3-, 10-, 12-, 14- and 15-bar surface pressures. For these sensitivity tests, no clouds were incorporated. For atmospheric pressures greater than 1 bar, the vertical pressure profile was multiplied by the desired integer scaling factor, and the molecular weight of the atmosphere was adjusted accordingly. We did two runs at each atmospheric concentration: One with the actual fine-grained snow albedo spectrum from Grenfell (\citeyear{Grenfell1994}), which drops to low values in the near-IR, and one where the Grenfell snow spectrum was altered to exhibit artificially high albedo values of 0.6 at wavelengths longer than 1.1 $\mu$m. Cross-section and line absorption coefficient files for CO$_2$ were included as input and created using LBLABC with data from the HITRAN 2008 database \citep{Rothman2009}. The results are plotted in Figure 18.

The pure nitrogen atmosphere run resulted in the highest broadband planetary albedo, due to its relative dearth of spectral absorption features. When 0.1 bar of atmospheric CO$_2$ is introduced, the broadband planetary albedo decreases  by 7\%. When the artificially-enhanced snow spectrum is used as the input surface spectrum, the broadband albedo values for the pure nitrogen atmosphere case increase by the largest amount. As atmospheric CO$_2$ concentration is increased, the difference between the calculated broadband planetary albedo using the actual snow spectrum, and that using the artificially higher snow spectrum decreases. At atmospheric concentrations of $\sim$3 bars of CO$_2$, the broadband albedo for the planet with the actual snow spectrum matches that calculated with the artificially-high snow spectrum to within 6.5\%. At 10 bars of CO$_2$, the two values match to within two-tenths of one percent. Differences in reflectivity as a function of wavelength between the planets with the actual and altered snow surfaces, given CO$_2$ concentrations of 0.1, 3, and 10 bars, are shown in Figure 19. With 10 bars of CO$_2$ in the atmosphere, the reflectivity at longer wavelengths has plummeted due to near-IR absorption by CO$_2$, and the spectra of the planets with the actual and artificially-enhanced snow albedo spectra are indistinguishable. 

The absorption cross-section of CO$_2$ is temperature-dependent, with cross-sections increasing with temperature towards longer wavelengths \citep{Parkinson2003}. Increasing the temperature at which our absorption coefficients were calculated in SMART by 50 K for atmospheres of 10 bars of CO$_2$ resulted in a $\sim$10\% decrease in broadband planetary albedo. There is therefore more work to do in the future on the temperature-dependent CO$_2$ absorption profile and its influence on the snow and ice-albedo effect on M-dwarf planets.

As GCMs are currently incapable of simulating atmospheric CO$_2$ concentrations greater than $\sim$0.1 bar, we used SMART with our 1-D EBM to address the high-CO$_2$ case. Figure 20 shows the results of EBM simulations using broadband planetary albedos calculated with SMART assuming a 3-bar pure CO$_2$ atmosphere on planets orbiting M-, G-, and F-dwarf stars, along with the original EBM simulations with Earth-like atmospheric CO$_2$ concentrations from Figure 6. Inputs to SMART, including the atmospheric pressure profile and molecular weight of the atmosphere, were scaled for a 3-bar surface pressure. As in the model runs with an Earth-like atmosphere, broadband planetary albedos for different surface types were generated assuming both clear and cloudy sky conditions. Average broadband planetary albedos over ocean, land, and below-freezing surfaces for the high-CO$_2$ case were calculated as described in Section 3, and used as input to the EBM. The coefficients governing the effect of CO$_{2}$ and water vapor in the linear function of temperature that is parameterized for outgoing longwave radiation in the EBM were also adjusted to account for the increased CO$_2$. Due to near-IR absorption by the significantly larger amount of CO$_2$ in the atmospheres of the three planets, ice-free climates exist on all planets with much higher decreases in instellation than in the simulations with an Earth-like CO$_2$ concentration. Once ice appears on each planet, the ice expands to total ice cover with a smaller reduction in instellation than with Earth-like CO$_2$ levels, as the planets are close to the OHZ at these low values of stellar flux density. This increased climate sensitivity is the result of much lower surface temperatures at those low values of instellation, as well as the weaker relationship between outgoing longwave radiation and surface temperature in this high-CO$_2$ case. In essence, the greenhouse effect fails to keep the surface temperature above freezing over large areas of the planet at these low values of instellation as we approach the maximum CO$_2$ greenhouse limit. Figure 20 indicates that given high levels of CO$_2$, planets remain ice-free throughout the majority of the width of the habitable zone. M-dwarf planets with 3-bars of CO$_2$ in their atmospheres appear ice-free with larger decreases in instellation than G- or F-dwarf planets with equivalent CO$_2$ concentrations. This is due to greater CO$_2$ absorption of the increased near-IR radiation emitted by M-dwarf stars. M-dwarf planets appear to be less susceptible to global glaciation over the course of their history, but at the outer edge of the habitable zone, where CO$_2$ levels are expected to be high, the ice-albedo effect is muted by atmospheric CO$_2$ absorbing strongly in the near-IR. Once ice does appear on the planets, even with the largest-grained, lowest-albedo ice surfaces, the points of global ice cover do not occur farther out than the traditional maximum CO$_2$ greenhouse limits \citep{Kasting1993}, which have been updated by Kopparapu \emph{et al.} (\citeyear{Kopparapu2013}) and we have plotted as vertical lines for each planet in Figure 20.

\section{Discussion}

The results of our model simulations indicate that the interaction between wavelength-dependent planetary surface albedo and the SED of a star may significantly affect the climate of an orbiting planet. The spectral dependence of ice and snow albedo is most important for planets in the middle range of a star's habitable zone. At the inner edge of the habitable zone, a planet will receive too much flux to have significant ice on its surface. At the outer edge of the habitable zone, high CO$_2$ masks the ice-albedo effect. 

Our SMART + EBM model can reproduce the Earth's current global mean surface temperature and ice line latitude at its present obliquity to within three degrees Celsius and six degrees of latitude respectively, given gas absorption by the seven major atmospheric constituent species. EBM simulations produce ice lines that are farther towards the poles on M-dwarf planets compared to those on G-dwarf planets for a given instellation, providing confirmation of the analytical supposition of Joshi and Haberle (\citeyear{Joshi2012}) that the cooling effect of ice-albedo feedback may indeed be suppressed on planets orbiting stars that emit large amounts of radiation in the near-IR, where ice and snow absorb strongly. Building on their initial work, our EBM results additionally imply that the cooling effect of ice-albedo feedback may be enhanced on planets orbiting F-dwarf stars, as global ice cover occurs for a much smaller change in instellation on F-dwarf planets than on planets orbiting G- or M-dwarf stars (Figs. 6 and 8). SMART upwelling and downwelling flux output indicate that broadband planetary albedo increases for SEDs with higher visible and near-UV radiation output, especially for ice and snow surface spectra, whose wavelength-dependent albedo increases in the visible and near-UV. This trend remains even when the effect of Rayleigh scattering in the atmosphere is removed, and would be absent without incorporating the spectral dependence of ice and snow albedo across the full near-UV through near-IR spectrum.  Planets orbiting cooler, redder stars remain free from global ice cover with larger decreases in instellation.

Additionally, we find that ice grain size strongly affects the latitudinal extent of the ice line, and that the type of ice on a planetary surface may strongly affect its climate. The larger the ice grain size, the lower the calculated broadband planetary albedo, and the higher the resulting ice line latitude determined by the EBM. Given a simplified aqua planet configuration, ice-covered states are possible with smaller decreases in instellation for below-freezing surfaces of finer grain size such as fresh snow, which have high broadband albedos given an F-dwarf incident spectrum. Assuming an Earth-like atmosphere with clouds and the effect of Rayleigh scattering included, our SMART runs show that increasing the ice grain size results in a $\sim$12\% to $\sim$17\% increase in broadband planetary albedo per 25\% mixture of blue ice and snow. Planets with surfaces composed of ice of larger grain size appear to be more stable against snowball episodes.

To further explore the atmospheric response to instellation of varying SED, we ran 3-D GCM simulations as well. In our GCM runs we have assumed an aqua planet with a 50-meter slab ocean.  With a 10\% reduction in instellation, surface temperatures are warmer on an orbiting M-dwarf aqua planet than a G-dwarf aqua planet, due to greater atmospheric absorption and emission of the large proportion of near-IR radiation from M-dwarf stars, by greenhouse gases like CO$_2$ and water vapor, which have many absorption bands in the near-IR and IR regions of the spectrum. Therefore, atmospheric absorption and downwelling longwave heating play a large role in reducing the fraction of ice covering M-dwarf planets. 

However, the role played by the spectral dependence of ice albedo in affecting climate is not insignificant. In our GCM simulations of M-dwarf planets receiving far less instellation than G-dwarf planets, we find that a higher percentage of absorption by surface ice of the reduced amount of incident shortwave radiation on the M-dwarf planet helps to compensate for the reduced instellation. Lowering the albedos of ice and snow decreased the climate sensitivity of the M-dwarf planet compared to that using the default albedo values. This result implies that increased shortwave absorption by surface ice increases climate stability to changes in instellation, and planetary climate appears to be affected by the spectral dependence of ice albedo, even in the presence of Earth-like greenhouse gases in the atmosphere. 

The slopes in Figure 8 are a measure of the climate sensitivity to changes in instellation for these planets, given their host star SEDs and the concentration of CO$_2$ held fixed at the present atmospheric level on Earth. They indicate a larger change in ice cover for a given change in instellation for planets with incident spectra like the Sun's, or bluer, compared to redder host stars. The G-dwarf planet has entirely frozen over at 92\% of the modern solar constant. This result is similar to those of previous climate simulations carried out by Abe \emph{et al.} (\citeyear{Abe2011}), who find that an aqua planet around the Sun becomes ice-covered with a 10\% reduction in instellation. A snowball state is not reached in our GCM simulations of an M-dwarf planet until the amount of flux from the star has been reduced to 73\% of the modern solar constant. An F-dwarf planet has completely frozen over with only a 2\% reduction in instellation compared to that of present Earth. M-dwarf planets therefore appear much less sensitive to changes in instellation, and thus less susceptible to snowball states. The Neoproterozoic Snowball Earth episodes of $\sim$750 to $\sim$635 Myr ago have been linked to the emergence of multicellular life on Earth, by enhancing the flux of dissolved phosphates into the ocean, causing increased primary productivity and organic carbon burial, and leading to the rise of oxygen in the ocean and atmosphere \citep{Planavsky2010}. Planets less likely to experience such global-scale glaciations may therefore be dependent on alternate pathways to serve as catalysts for biological evolution. The M-dwarf planet in our simulations also exhibited stable lower latitude ice lines than the G- or F-dwarf planets, and this may be due to the lower-albedo ice formed on its surface. A more stable low-latitude ice line on M-dwarf planets may be possible as the result of a lower albedo contrast between bare sea ice and snow-covered ice \citep{Abbot2011}.

Land planets orbiting Sun-like stars, given their lower thermal inertia and drier atmospheres, completely freeze with a 23\% reduction in instellation, to 77\% of the modern solar constant \citep{Abe2011}. Given that M-dwarf aqua planets continue to exhibit ice-free areas until their stellar flux has been reduced by 27\% (to 73\% of the modern solar constant), M-dwarf planets may be more resistant to freezing than G-dwarf planets, regardless of land percentage. 

\subsection{Additional factors affecting planetary climate}
We have assumed a rotation period equal to the present Earth (24 hours) for our simulations, to isolate the effect of stellar SED and ice-albedo feedback on planetary climate. Synchronous rotation, which is expected to occur on M-dwarf planets orbiting in their stars' habitable zones (\citealp{Dole1964, Kasting1993, Joshi1997, Edson2011}), will certainly affect the results presented here. Earlier work by Edson \emph{et al.} (\citeyear{Edson2011}) indicates that increasing the rotation period of an aqua planet weakens low-latitude zonal winds and cools the planet, and increased atmospheric latent and oceanic diffusive heat transport could reduce temperature constrasts between the sunlit and dark sides of the planet. Given that a solar-type spectrum was used in their work, and our work here with an M-dwarf spectrum results in greater surface temperatures on M-dwarf planets than on G-dwarf planets receiving equal amounts of instellation, temperature contrasts between the sunlit and dark sides of a synchronously-rotating aqua planet orbiting an M-dwarf star at a 1-AU equivalent flux distance may be even further reduced.

Previous work using GCM simulations of synchronously rotating planets at 1-AU equivalent flux distances from a solar-type star suggests that atmospheric CO$_2$  concentrations could be nearly 200 times higher than the present atmospheric level on Earth if the substellar point is over a predominantly ocean-covered area, due to a lack of continental weathering \citep{Edson2012}. However, atmospheric CO$_2$ concentration could be limited by seafloor weathering on an aqua planet, resulting in much lower CO$_2$ partial pressures \citep{Edson2012}. Here, we have simulated an M-dwarf aqua planet with a 24-hour rotation period, making an Earth-like atmospheric CO$_2$ concentration a reasonable assumption. 

Although M-dwarf planet climates appear to be sensitive to the surface at present atmospheric levels of CO$_2$, there is a limit to the amount of CO$_2$ that can exist in the atmosphere of a planet without masking the surface entirely at near-IR wavelengths. While the SED of M-dwarf stars coupled with the spectral dependence of ice and snow albedo may indeed allow M-dwarf planets to exist ``Snowball free" with larger decreases in instellation than planets orbiting stars with higher visible and near-UV output, assuming low-to-moderate atmospheric levels of CO$_2$, CO$_2$ can be expected to build up in the atmosphere in response to lower surface temperatures and decreased silicate weathering \citep{Walker1981}. Abbot \emph{et al.} (\citeyear{Abbot2012}) found that provided the land fraction is at least $\sim$0.01, climate weathering feedback should operate without a strong dependence on land fraction. As our EBM simulations with high-CO$_2$ atmospheres used a land fraction of 0.01 (Section 2.2), then assuming a high-CO$_2$ atmosphere that could result from climate weathering feedback at the OHZ is not unreasonable. 

The high-CO$_2$ atmospheres ($>$3-10 bars) we have simulated using SMART may be sufficient to mask the suppressed ice-albedo feedback that is expected to occur on M-dwarf planets. From our EBM simulations of planets with 3-bar pure CO$_2$ atmospheres, the increased absorption of near-IR radiation by large amounts of CO$_2$ prevents ice from forming on all three planets with significantly larger decreases in instellation than in the simulations with Earth-like CO$_2$ levels. The increased climate sensitivity that occurs once ice does appear on a planet with 3 bars of CO$_2$ in its atmosphere is the result of the greenhouse effect failing to keep surface temperatures above freezing at low values of instellation near the OHZ. We would expect to see this behavior persist in GCM simulations as well. Climate model simulations of Earth-like planets at the outer edge of their stars' habitable zones, with CO$_2$ partial pressures of 3-10 bars, are outside of the realm of present GCM capabilities. 

Higher concentrations of CO$_2$ appear to lower the calculated broadband albedo of planets with M-dwarf host stars  (though the amount of reduced albedo decreases as CO$_2$ increases, as indicated by the flattening of the curve in Figure 18, bottom). For M-dwarf host stars, this may be largely due to increased CO$_2$ atmospheric absorption in the near-IR. However, increased Rayleigh scattering due to high-CO$_2$ concentrations is expected to lead to increases in planetary albedo, at least for G-dwarf host stars such as the Sun \citep{Kasting1986}. The effect of Rayleigh scattering could dominate atmospheric absorption at higher pressures. However, given that the effect of Rayleigh scattering is more pronounced for shorter-wavelength incident radiation,  our decreased planetary albedos for higher-pressure atmospheres may be the consequence of atmospheric absorption masking high-albedo snow, given that Rayleigh scattering is less important on planets receiving higher percentages of long-wavelength radiation from their host stars (\citealp{Kasting1993, Pierrehumbert2011a, Kopparapu2013}). There may therefore be a particular stellar mass (somewhere between AD Leo and the Sun) above which the combined effects of Rayleigh scattering and atmospheric near-IR absorption are maximized. Indeed, we found that the broadband albedo of a K-dwarf planet also decreased with higher concentrations of CO$_2$, and by a smaller percentage than that of our M-dwarf planet. Here we have assumed a 1-AU equivalent flux distance for the M-dwarf planet, so the likelihood of CO$_2$ cloud formation in the atmosphere is low. Previous work with a 1-D radiative-convective model has suggested that CO$_2$ ice clouds in the atmosphere may result in net warming of a planet's surface \citep{Forget1997}.  

Given that the silicate weathering thermostat adjusts the CO$_2$ concentration to increase with decreasing surface temperature, and M-dwarf planets appear less sensitive to changes in instellation (as indicated by the shallower change in temperature and ice extent on the M-dwarf planet shown in Figures 6 and 8), habitable M-dwarf planets can be expected to exhibit lower concentrations of CO$_2$ at a given distance from their host stars than planets orbiting stars with higher visible and near-UV output, essentially requiring lower amounts of CO$_2$ in order to remain free of global ice-cover at lower values of instellation. We have not explored the effect of decreased CO$_2$ on the percent of the modern solar constant required for global ice cover on an M-dwarf planet using a GCM. Given the increased atmospheric stability in the troposphere of the M-dwarf planet receiving 73\% of the modern solar constant compared to the G-dwarf planet receiving 92\% of the modern solar constant (Figure 15), we expect the M-dwarf planet to be more resistant to global glaciation with decreased CO$_2$ than a G-dwarf planet, given an equivalent reduction in atmospheric CO$_2$ concentration. The difference between M-dwarf planets and planets orbiting hotter main-sequence stars becomes less noticeable at the outer edge of the habitable zone (where the masking effect of high CO$_2$ becomes more important) than closer in to the star (where CO$_2$ levels would be lower, and more ice would form on a planet's surface). 

Our EBM simulations with 3-bar CO$_2$ atmospheres indicate that the spectral dependence of surface ice albedo is less important at the outer edge of the habitable zone, as planets remain free of planetary ice cover throughout the majority of the width of the habitable zone, due to large amounts of near-IR absorption by CO$_2$. We find that M-dwarf planets become ice-covered at $\sim$33\% of the modern solar constant with 3 bars of CO$_2$ in their atmospheres, given the lowest-albedo ice surface. This value, while much smaller than that required for ice-covered conditions given Earth-like CO$_2$ levels (73\% of the modern solar constant, from our GCM results), is still greater than the maximum CO$_2$ greenhouse limit, derived by Kasting \emph{et al.} (\citeyear{Kasting1993}) and updated by Kopparapu \emph{et al.} (\citeyear{Kopparapu2013}), of 25\% of the modern solar constant for a planet orbiting an M0 star. It has been postulated that the OHZ may be 10-30\% farther out from an M-dwarf star given the effect of the wavelength-dependent reflectivity of surface ice and snow on planetary albedo \citep{Joshi2012}. However, CO$_2$ partial pressures of 10 bars or more may be required to maintain atmospheric stability and surface liquid water at the outer edge of the habitable zone for M-dwarf stars \citep{Wordsworth2011}, and we have shown that high levels of atmospheric CO$_2$ entirely mask the climatic effect of surface water ice and snow. Therefore, we find the traditional outer edge of the habitable zone to be unaffected by the spectral dependence of ice and snow albedo. 

\section{Conclusions}

We have demonstrated using 1-D and 3-D climate simulations that planets orbiting  cooler, redder stars at equivalent flux distances exhibit higher global mean surface temperatures than planets orbiting stars with more visible and near-UV radiation output. The increased surface temperatures are in large part the consequence of absorption by atmospheric gases of the significantly higher percentage of near-IR radiation emitted by M-dwarf stars. However, we have shown that the spectral dependence of water ice and snow albedo does play a role in affecting climate. Changes to planetary climate appear to be less sensitive to M-dwarf SED than G- and F-dwarf SED, as evidenced by the smaller change in ice extent for a given change in stellar flux for M-dwarf planets. At a fixed level of CO$_2$, M-dwarf aqua planets remain free of global ice cover with 25\% less stellar flux than F-dwarf aqua planets, and 19\% less stellar flux than G-dwarf aqua planets. M-dwarf planets may therefore be more stable against low-latitude glaciation over the course of their history than planets orbiting stars with higher visible and near-UV radiation output. As low-latitude glaciation has been linked to the emergence of complex life on Earth, planets less predisposed to such glacial episodes would need to rely on other means to aid in biological evolution. M-dwarf planets may also exhibit more stable low-latitude ice lines as a consequence of lower-albedo ice on their surfaces. At the outer edge of the habitable zone, where CO$_2$ can be expected to increase with decreasing surface temperature and silicate weathering, the spectral dependence of surface ice and snow albedo is less important, and does not extend the traditional outer edge of the habitable zone given by the maximum CO$_2$ greenhouse. However, due to their lower climate sensitivity to changes in instellation, M-dwarf planets would likely have lower amounts of CO$_2$ in their atmospheres far out in their stars' habitable zones than planets orbiting stars with higher visible and near-UV output at an equivalent flux distance, and require less CO$_2$ in order to maintain clement conditions for surface liquid water.  

\section{Acknowledgments}

This material is based upon work supported by the National Science Foundation Graduate Research Fellowship Program under Grant Nos. DGE-0718124 and DGE-1256082. This work was performed as part of the NASA Astrobiology Institute's Virtual Planetary Laboratory Lead Team, supported by the National Aeronautics and Space Administration through the NASA Astrobiology Institute under Cooperative Agreement solicitation NNH05ZDA001C. The authors wish to thank two anonymous referees for their comments and suggestions, which greatly improved the paper. A.L.S. thanks David Crisp and Amit Misra for useful conversations about F-dwarf stars, ozone, and SMART, Eric Agol for help with stellar flux calculations, John Armstrong for helpful IDL stellar interpolation code, and John Johnson and Carl Grillmair for helpful advice regarding graphics. Special thanks to the Virtual Planet Laboratory for computer models and resources, the NASA Astrobiology Institute, the NSF Astrobiology IGERT program, Bruce Briegleb and Tom Ackerman. Daniel Koll provided assistance with running the GCM in aqua planet mode. Stephen Warren provided advice about spectral albedos of snow and ice.

\section{Author Disclosure Statement}
No competing financial interests exist.	
\newpage

\bibliography{/Users/aomawashields/Documents/research/thesis/paper1/Shields_refs.bib}

\linespread{1.0}
\begin{table}[!htp] 
\caption{CAM4 spectral wavelength bands specifying shortwave (stellar) incoming flux into the atmosphere, and the percentage of flux within each waveband for the Sun, M-dwarf star AD Leo, and F-dwarf star HD128167.} 
\vspace{2 mm}
\centering \begin{tabular}{c c c c c c} 
\hline\hline 
Band & $\lambda_{min}$ & $\lambda_{max}$ & Sun \% flux & AD Leo \% flux  & HD128167 \% flux\\  [0.5ex]
\hline
$1$ & 0.200 & 0.245 & 0.124 & 0.025 & 3.291 \\
$2$ & 0.245 & 0.265 & 0.130 & 0.009 & 0.717 \\
$3$ & 0.265 & 0.275 & 0.177 & 0.0029 & 0.597 \\
$4$ & 0.275 & 0.285 & 0.167 & 0.008 & 0.528 \\
$5$ & 0.285 & 0.295 & 0.349 & 0.003 & 0.763 \\
$6$ & 0.295 & 0.305 & 0.399 & 0.004 & 0.917 \\
$7$ & 0.305 & 0.350 & 2.805 & 0.029 & 4.620 \\
$8$ & 0.350 & 0.640 & 36.00 & 3.390 & 43.12 \\
$9$ & 0.640 & 0.700 & 6.643 & 1.895 & 6.220 \\
$10$ & 0.700 & 5.000 & 52.79 & 94.64 & 40.68 \\ 
$11$ & 2.630 & 2.860 & 0.613 & 1.783 & 0.393 \\ 
$12$ & 4.160 & 4.550 & 0.175 & 0.749 & 0.114 \\ [1ex]
\hline 
\end{tabular} 
\label{table:nonlin} 
\end{table}
\pagebreak
\begin{table}[!htp] 
\caption{Broadband planetary albedos calculated with upwelling and direct downwelling stellar flux outputs from SMART for fine-grained snow, large-grained blue marine ice, ice of intermediate density between the two end-members, ocean, and land surfaces, given the SEDs of F-dwarf star HD128167, the Sun (a G-dwarf star), K-dwarf star HD22049, and M-dwarf AD Leo. For the planetary albedos that include gases and clouds (right column), average broadband albedos were calculated as described in Zsom \emph{et al.} (\citeyear{Zsom2012}) assuming 64\% cloud cover \citep{Warren2002}, with 40\% low stratocumulus clouds and 24\% high cirrus clouds. Random overlap between the two cloud layers is assumed \citep{Oreopoulos2003}.} 
\vspace{2 mm}
\centering \begin{tabular}{c c c c c} 
\hline\hline 
Stellar Type & surface & No gases or clouds & No gases or clouds & Earth-like gases, clouds\\  [0.5ex]
& & Rayleigh scattering on & Rayleigh scattering off & Rayleigh scattering on\\
\hline
F-dwarf & snow & 0.84988 & 0.83361 & 0.66833 \\
& 75\% & 0.72756 & 0.70568 & 0.59884 \\
& 50\%  & 0.60744 & 0.57704 & 0.53664 \\
& 25\%  & 0.48993 & 0.44840 & 0.47961 \\
& blue ice & 0.37330 & 0.31782 & 0.42542 \\
& ocean & 0.19349 & 0.07221 & 0.32865 \\
& land & 0.40438 & 0.30470 & 0.41428 \\
\hline
G-dwarf & snow & 0.79724 & 0.79577 & 0.64622 \\
& 75\% &  0.66882 & 0.66421 & 0.57636 \\
& 50\% & 0.54191 & 0.53194 & 0.51363 \\
& 25\% & 0.41708 & 0.39966 & 0.45585 \\
& blue ice & 0.29312 & 0.26603 & 0.40093 \\
& ocean & 0.14155 & 0.07070 & 0.31948 \\
& land & 0.38829 & 0.33513 & 0.41484 \\
\hline
K-dwarf & snow & 0.74842 & 0.74784 & 0.60392 \\
& 75\% &  0.62041 & 0.61768 & 0.53716 \\
& 50\% & 0.49350 & 0.48685 & 0.47708 \\
& 25\% & 0.36822 & 0.35602 & 0.42150 \\
& blue ice & 0.24418 & 0.22435 & 0.36870 \\
& ocean & 0.11921 & 0.06965 & 0.30235 \\
& land & 0.38606 & 0.35098 & 0.40133 \\ 
\hline
M-dwarf & snow & 0.51963 & 0.51905 & 0.39800 \\
& 75\% & 0.41245 & 0.41125 & 0.35478 \\
& 50\% & 0.30542 & 0.30308 & 0.31546 \\
& 25\% & 0.19890 & 0.19491 & 0.28406 \\
& blue ice & 0.09361 & 0.08749 & 0.24317 \\
& ocean & 0.07390 & 0.06518 & 0.23372 \\
& land & 0.37974 & 0.37516 & 0.33165 \\ [1ex]
\hline
\end{tabular} 
\label{table:nonlin} 
\end{table}
\pagebreak

\linespread{1.15}

\begin{figure}[!htb]
\begin{center}
\includegraphics [scale=2.00]{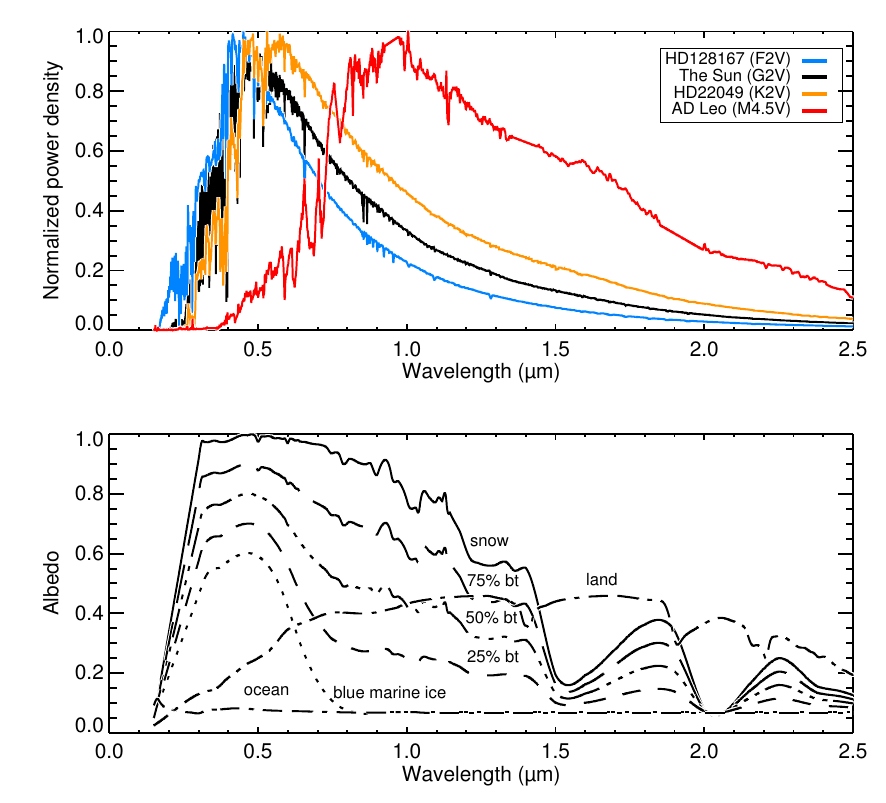}\\
\caption{Top: the SEDs for F-, G-, K-, and M-dwarf stars, normalized by the peak flux. Bottom:  the spectral distribution of fine-grained snow, blue marine ice, and 25\%, 50\%, and 75\% mixtures of the two end-members. Ocean and land spectral distributions used in the radiative transfer and energy balance models are also plotted.}
\label{Figure 1. }
\end{center}
\end{figure}

\begin{figure}[!htb]
\centering
\includegraphics [scale=2.00]{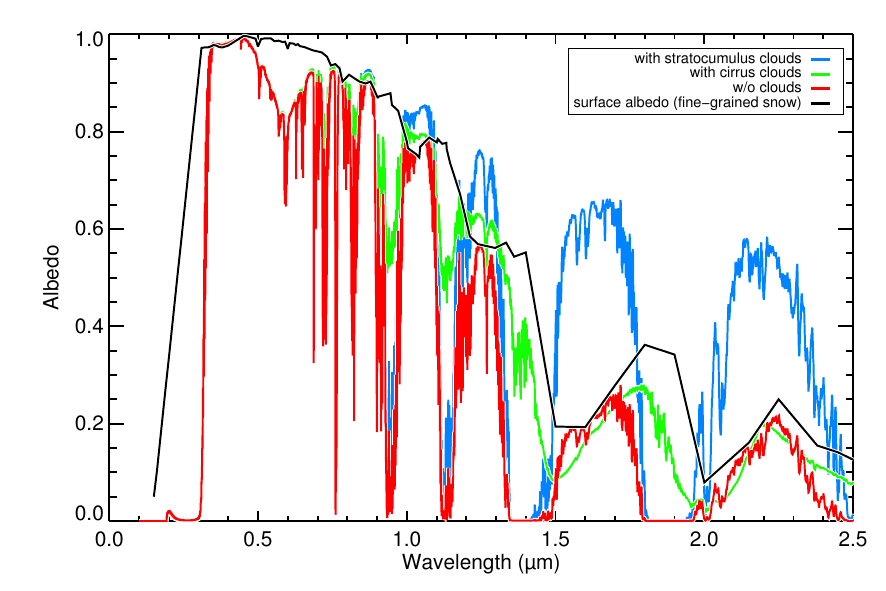}
\caption{Wavelength-dependent reflectivity of a planet with an Earth-like atmosphere, an underlying snow surface, clearsky conditions (red), 100\% cirrus cloud cover (green), and 100\% stratocumulus cloud cover (blue), calculated using SMART. The empirical spectrum for fine-grained snow (from Figure 1) is plotted here (black) for reference.}
\label{Figure.2.}
\end{figure}

\begin{figure}[!htb]
\begin{center}
\includegraphics [scale=1.00]{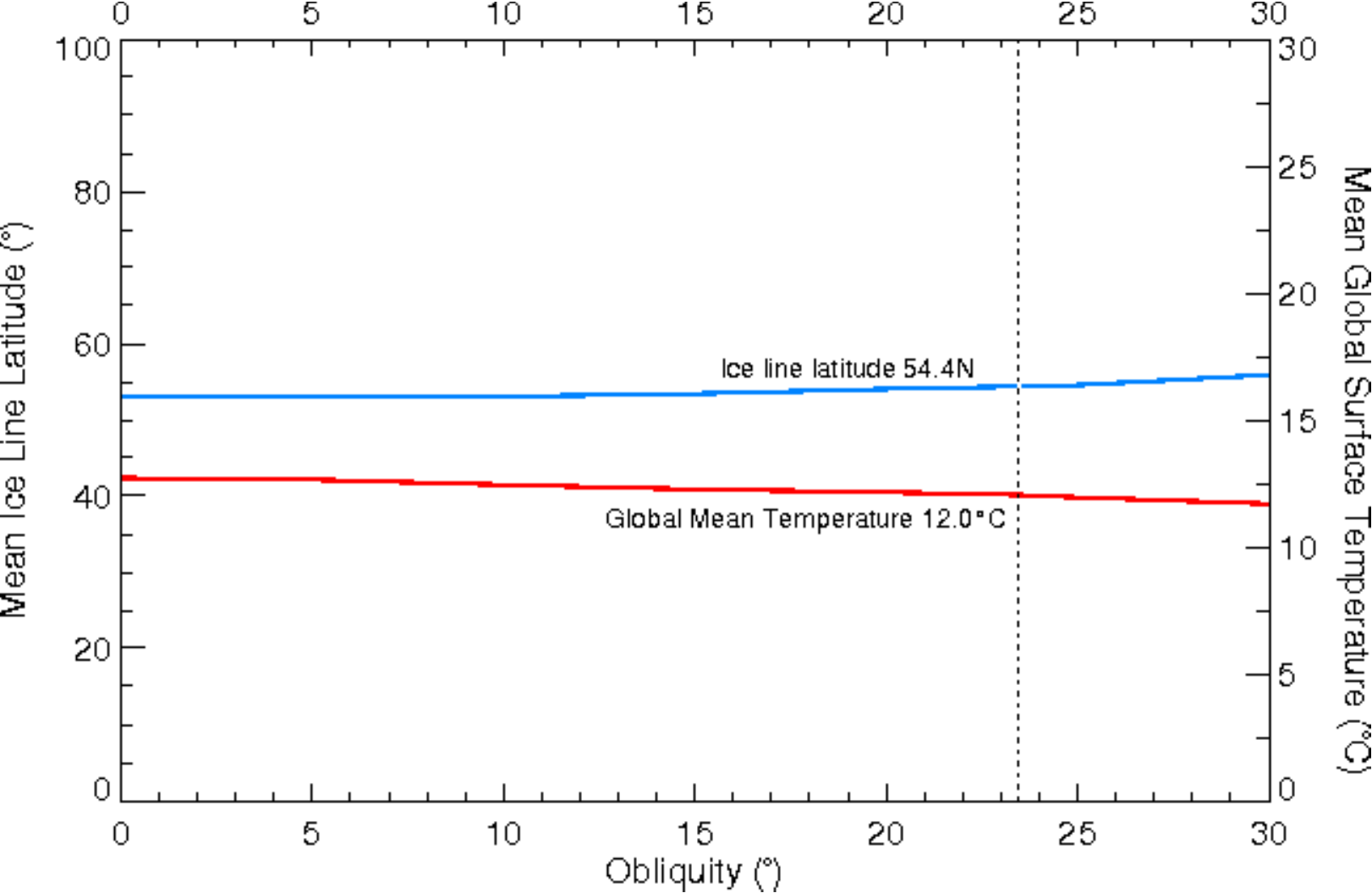}\\
\caption{Mean ice line latitude (blue) and global surface temperature (red) as a function of obliquity, calculated using a seasonal EBM. The Earth's northern hemisphere ice line latitude and global mean surface temperature at its present obliquity of 23.5$^\circ$C (vertical dashed line) is verified to within six degrees in latitude and three degrees Celsius, respectively. Ocean and land surfaces were assigned broadband albedos of 0.32 and 0.41, respectively (including atmosphere, for 36\% clear sky, 40\% stratocumulus cloud cover, and 24\% cirrus cloud cover). Regions where the surface temperature fell below -2$^\circ$C were assigned a broadband albedo of  0.46, which was calculated using the same percentages of clear sky and cloud cover, with the spectrum corresponding to the 25\% mixture of blue marine ice and fine-grained snow in Figure 1. Random overlap between the two cloud layers is assumed.}
\label{Figure 3.}
\end{center}
\end{figure}

\begin{figure}[!htb]
\begin{center}
\includegraphics [scale=2.00]{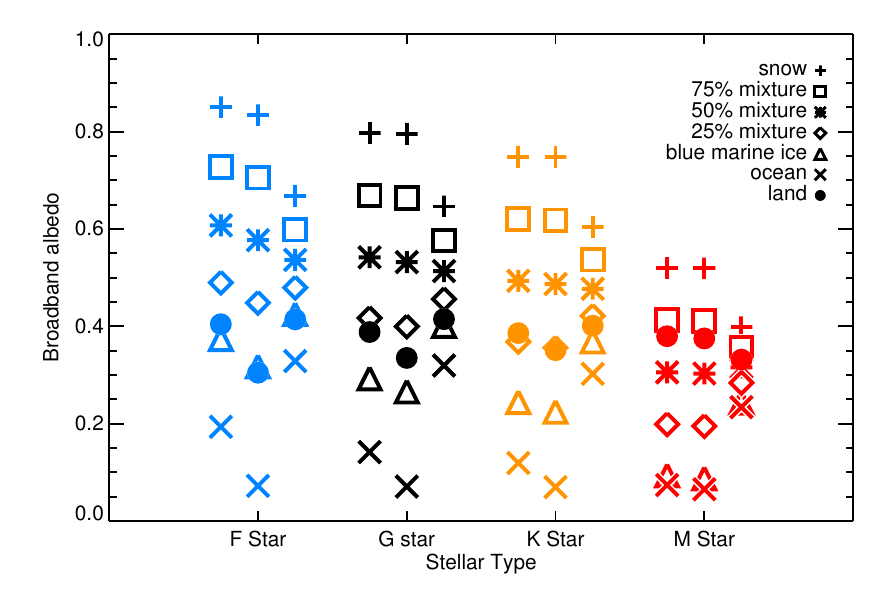}\\
\caption{Broadband planetary albedos calculated with upwelling and direct downwelling stellar flux outputs from SMART for ice, snow, ocean, and land surfaces given the SEDs of F-, G-, K-, and M-dwarf stars. For each star: Left -- no atmospheric gas absorption, but Rayleigh scattering is included; middle -- no gases or Rayleigh scattering (broadband F-dwarf planetary albedos for ice and snow surfaces are still larger than those for G-, K-, or M-dwarf planets, even after the effects of Rayleigh scattering are removed); right -- Rayleigh scattering, gas absorption, and clouds are included. These values are listed in Table 2. }
\label{Figure 4.}
\end{center}
\end{figure}

\begin{figure}[!htb]
\begin{center}
\includegraphics [scale=2.00]{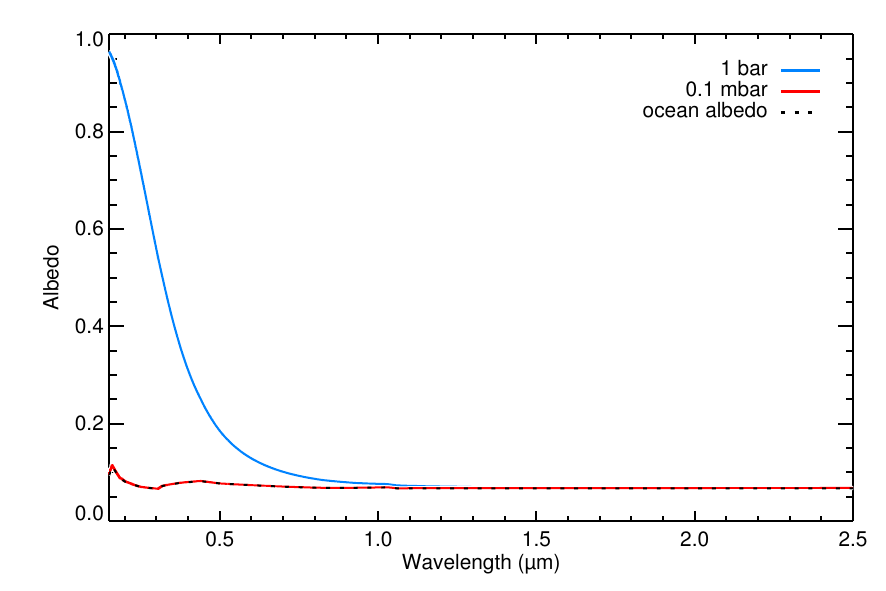}\\
\caption{Top-of-the-atmosphere (TOA) upwelling flux divided by the downwelling stellar flux (which is a measure of the planetary albedo) as a function of wavelength, for an ocean-covered planet with a surface pressure of 1 bar (blue), and 0.1 mbar (red) orbiting an F-dwarf star at an equivalent flux distance to the Earth around the Sun, calculated using SMART. No atmospheric gas absorption is included here. The rise in TOA flux (and therefore planetary albedo) evident in the 1-bar atmosphere case at $\lambda <$ 0.7 $\mu$m is due to Rayleigh scattering. At 0.1 mb, the Rayleigh scattering tail is absent, and matches the empirical albedo spectrum of the ocean surface from Figure 1 (dotted black).}
\label{Figure 5.}
\end{center}
\end{figure}

\begin{figure}[!htb]
\begin{center}
\includegraphics [scale=0.68]{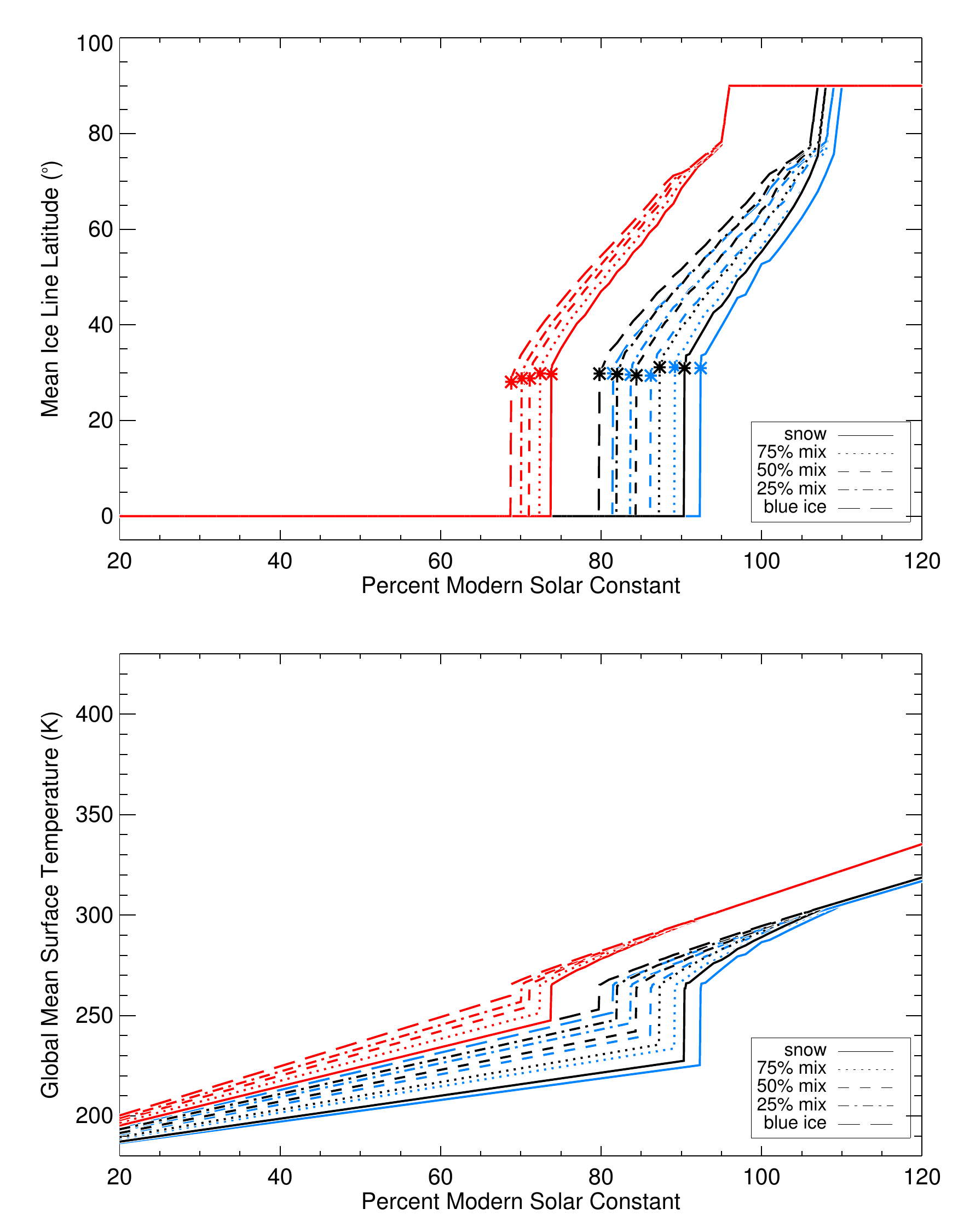}\\	
\caption{Mean ice line latitude (top) and global mean surface temperature (bottom) in the northern hemisphere as a function of percent of modern solar constant are calculated using a seasonal EBM, at present Earth obliquity (23.5$^\circ$) for aqua planets (land and ocean fraction 0.01 and 0.99, respectively) orbiting F-, G-, and M-dwarf stars at an equivalent flux distance. Below-freezing surfaces encountered during the EBM runs (where the temperature is less than -2$^\circ$C) were given broadband planetary albedos for ice and snow of varying grain size calculated from SMART at 1-bar surface pressure. EBM simulations were run with an initial warm start, with an approximate Earth-like zonal mean temperature distribution. The results  can be assumed to be similar for the southern hemisphere, given an assumed eccentricity of zero. The present atmospheric level (PAL) of CO$_2$ was used (F-dwarf planet in blue, G-dwarf planet in black, and M-dwarf planet in red). Asterisks denote the minimum ice line latitude before collapse to the equator and global ice coverage.}
\label{Figure 6.}
\end{center}
\end{figure}

\clearpage

\begin{figure}[!htb]
\begin{center}
\includegraphics [scale=0.40]{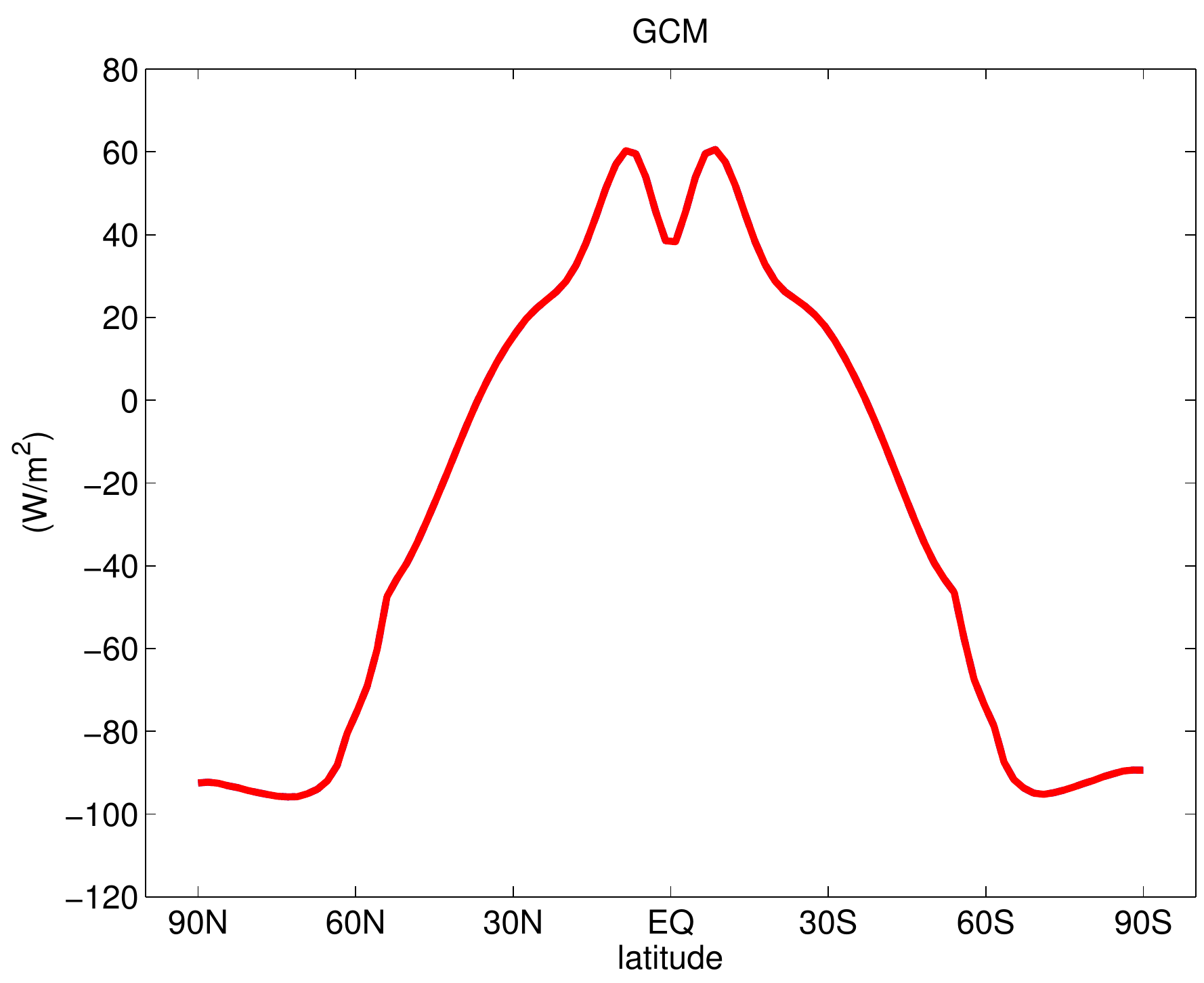}
\includegraphics [scale=0.40]{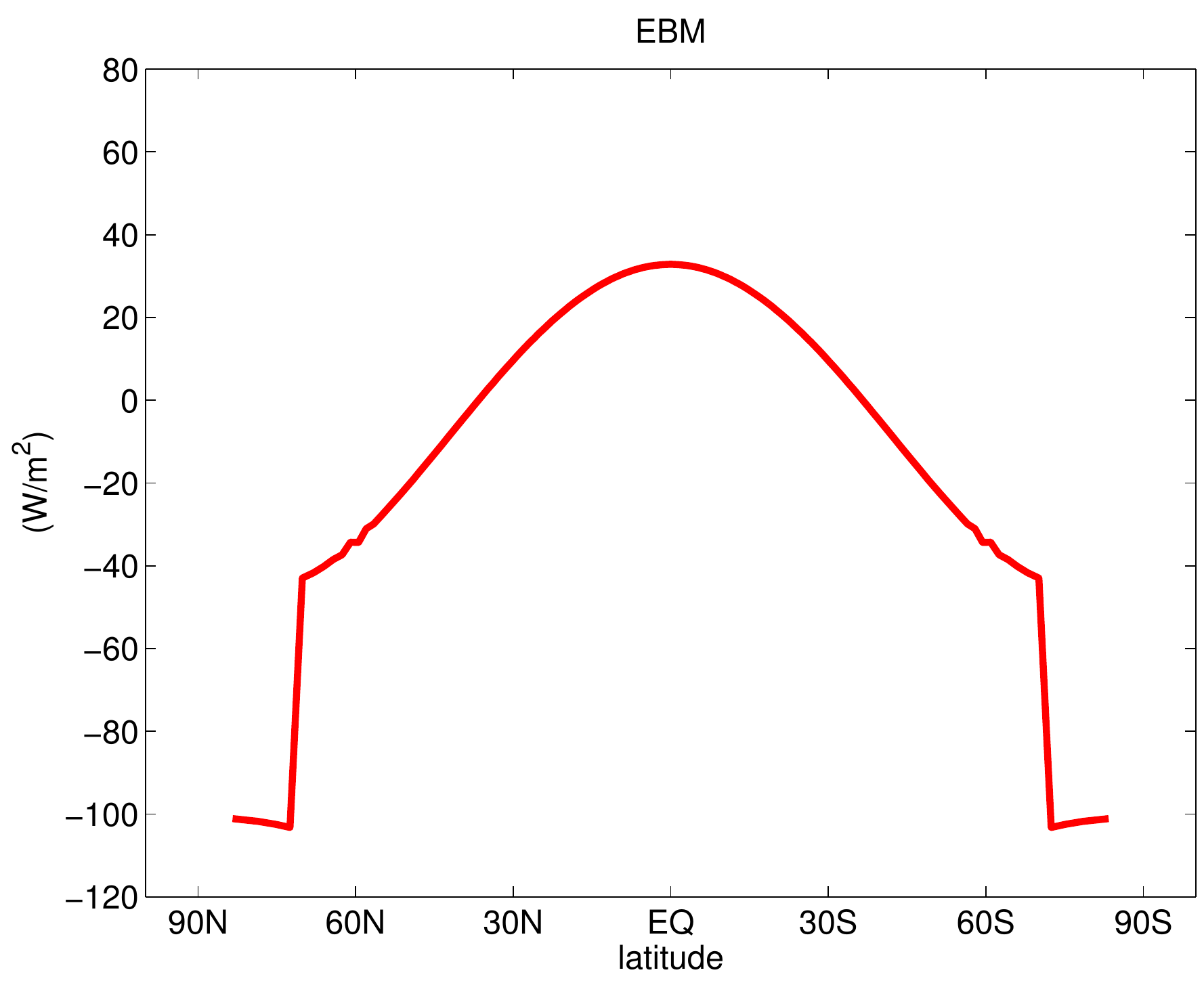}
\caption{Top-of-atmosphere absorbed shortwave radiation minus outgoing longwave radiation as a function of latitude, calculated in the GCM (left) and the EBM (right) for a G-dwarf planet receiving 100\% of the modern solar constant. When averaged over a few years or more, the net incoming heat flux must be equal to the divergence of heat from each grid cell in order to yield a net surface flux of zero locally and globally in our slab ocean model. The EBM shows a large jump in the net incoming heat flux near the poles, due to the abrupt change in albedo for ice-covered areas in the EBM, and lack of parameterized clouds (beyond our SMART treatment). A smoother transition in net incoming heat flux as a function of latitude is visible in the GCM, due to the presence of full-scale atmospheric dynamics, including clouds.}
\label{Figure 7.}
\end{center}
\end{figure}

\begin{figure}[!htb]
\begin{center}
\includegraphics [scale=1.75]{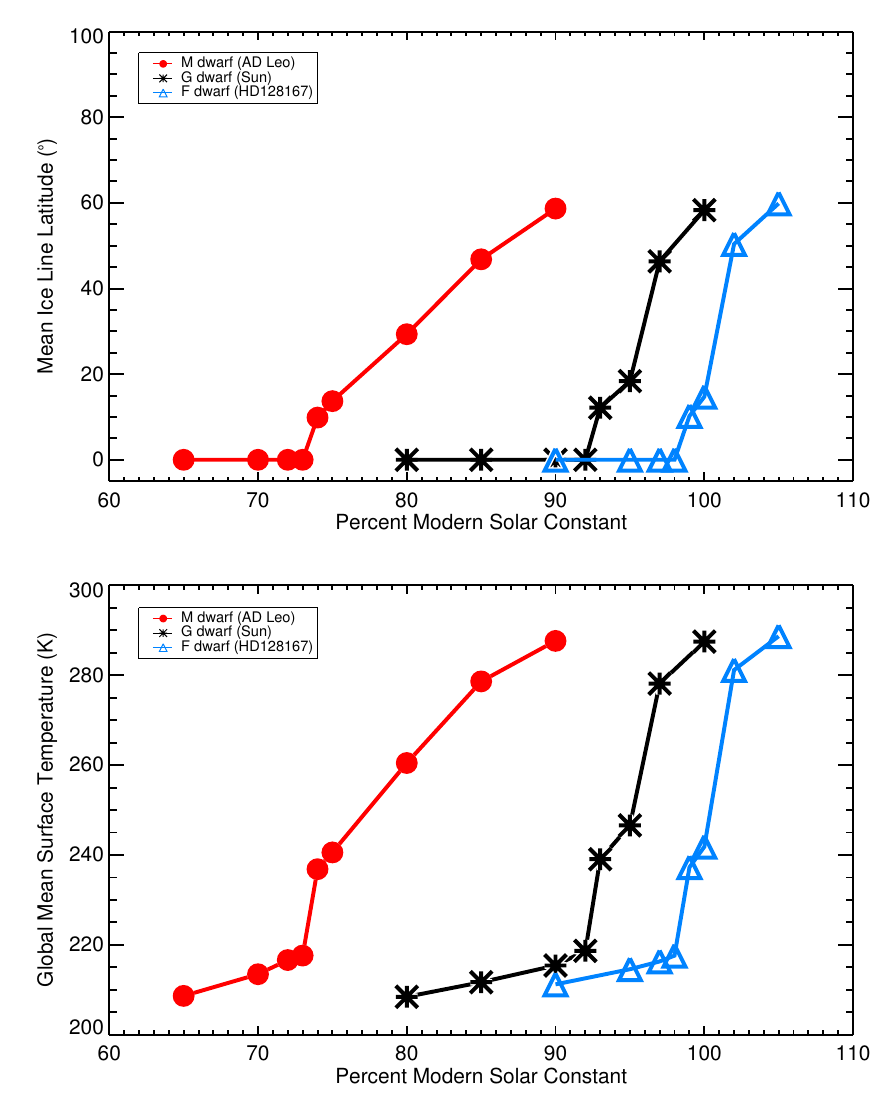}\\
\caption{Mean ice line latitude (top) and global mean surface temperature (bottom) as a function of percent of modern solar constant after a 40-year GCM run for an aqua planet orbiting the Sun (black), M-dwarf star AD Leo (red), and F-dwarf star HD128167 (blue) at equivalent flux distances. The slope of each line is a measurement of the climate sensitivity of the planet to changes in stellar flux. The shallower slope of the M-dwarf planet indicates a smaller change in surface temperature and ice extent for a given change in instellation than on the planets orbiting stars with greater visible and near-UV output.}
\label{Figure 8.}
\end{center}
\end{figure}

\begin{figure}[!htb]
\begin{center}
\includegraphics[scale=0.75]{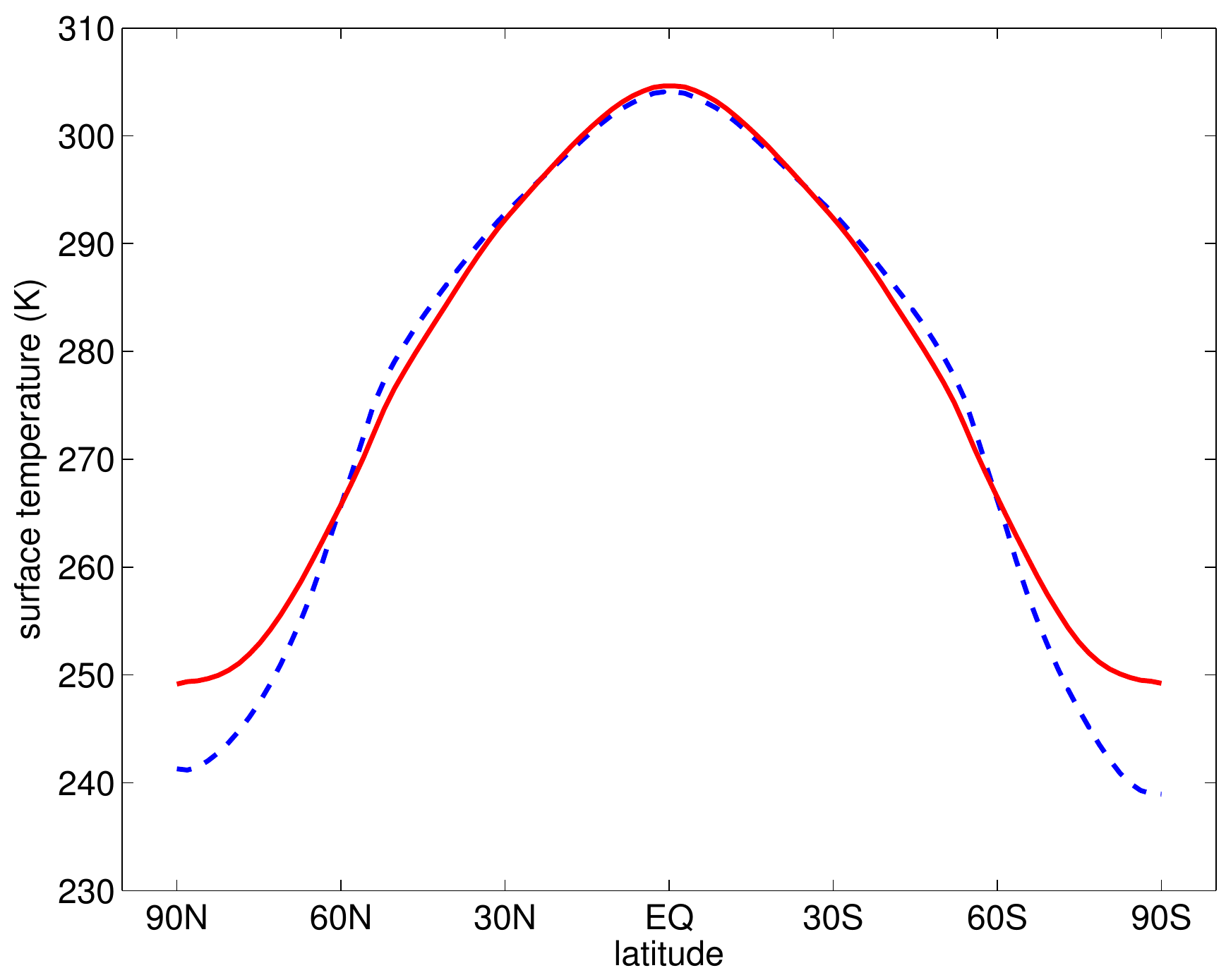}
\caption{Surface temperature on an aqua planet receiving 90\% of the modern solar constant from an M-dwarf star (red) compared with an aqua planet receiving 100\% of the modern solar constant from the Sun, a G-dwarf star (blue), after a 40-year GCM run.}
\label{Figure 9.}
\end{center}
\end{figure}

\begin{figure}[!htb]
\begin{center}
\includegraphics[scale=0.75]{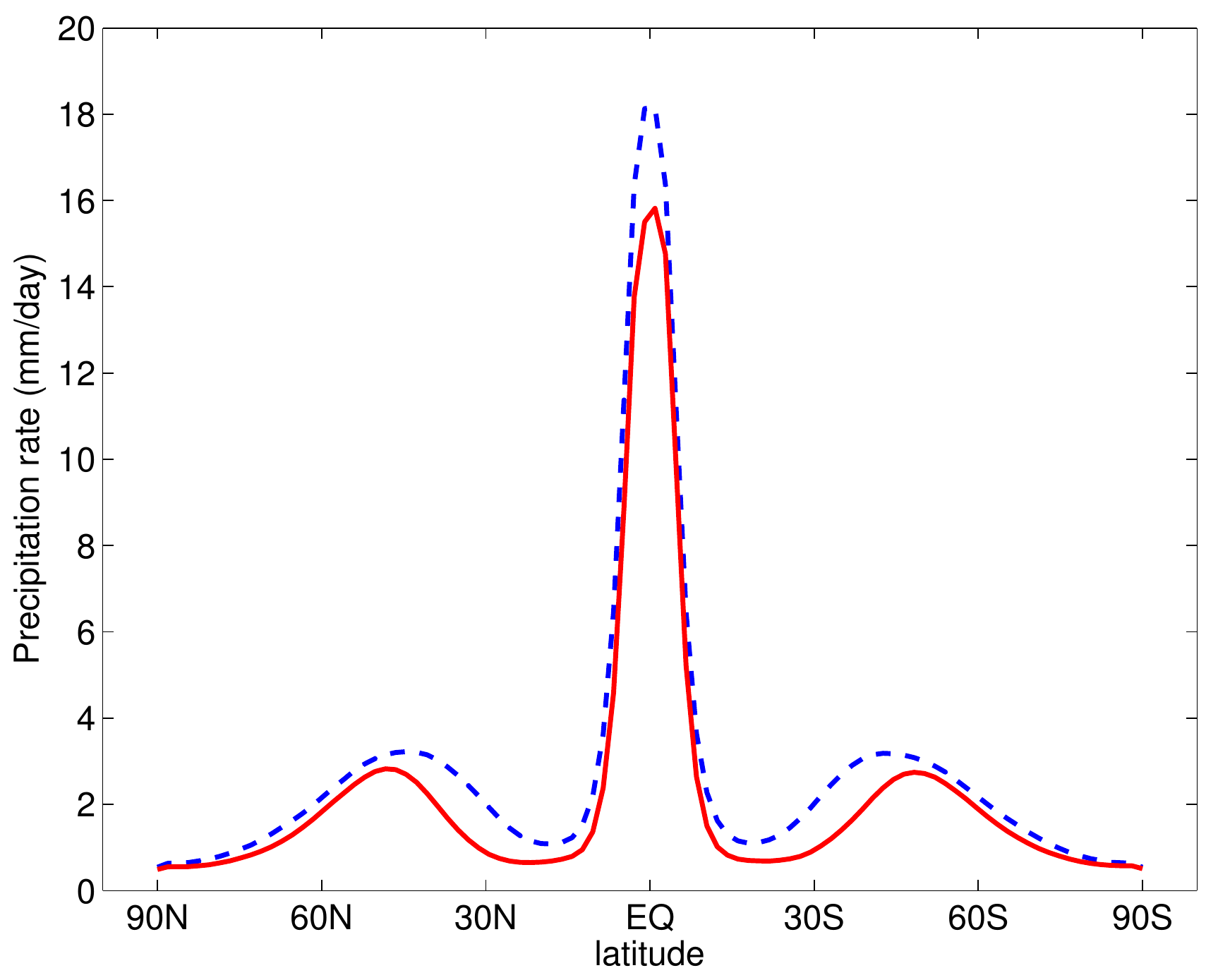}
\caption{Precipitation rate on an aqua planet receiving 90\% of the modern solar constant from an M-dwarf star (red) compared with an aqua planet receiving 100\% of the modern solar constant from the Sun (blue), after a 40-year GCM run.}
\label{Figure 10.}
\end{center}
\end{figure}

\begin{figure}[!htb]
\begin{center}
\includegraphics[scale=0.30]{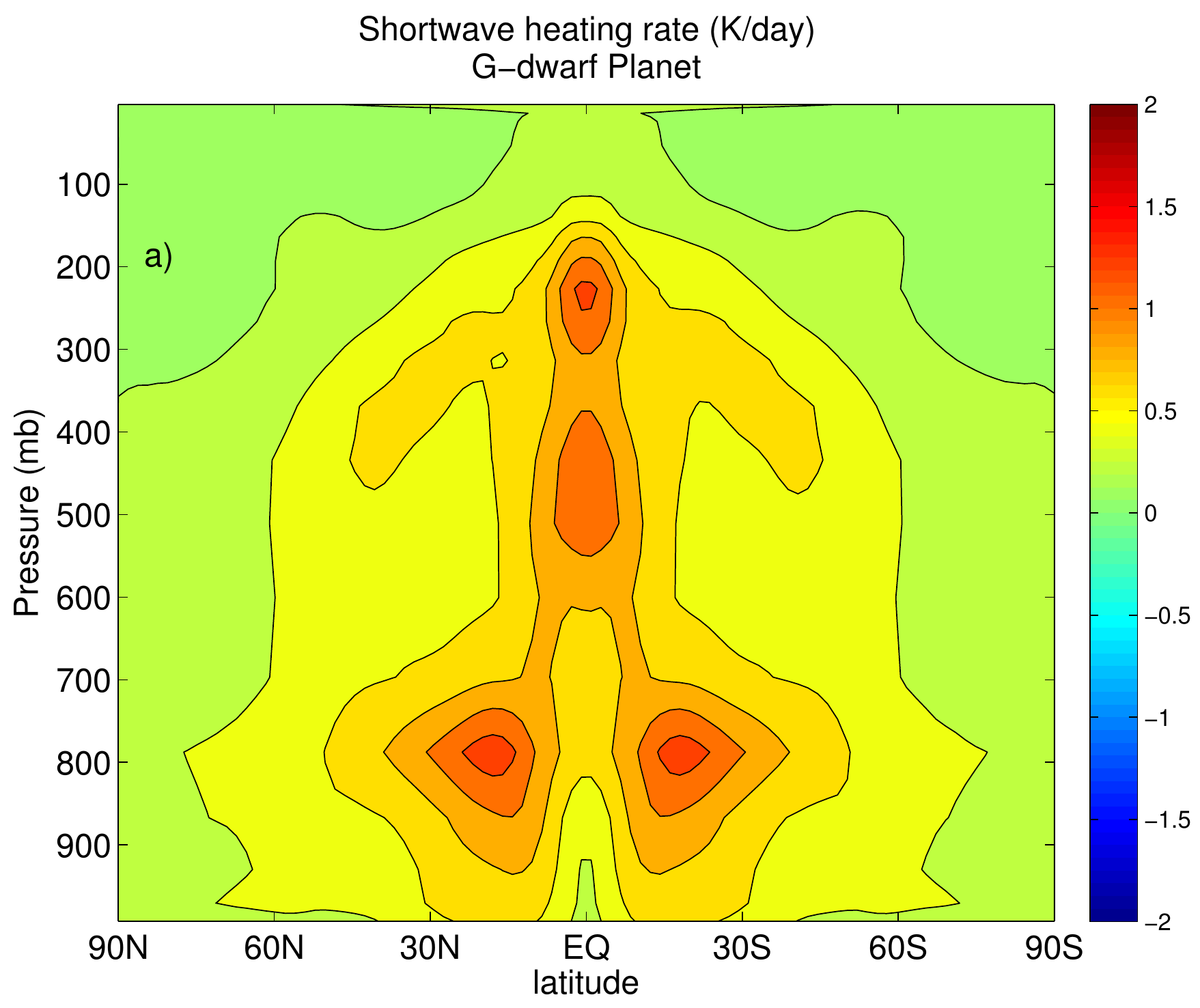}
\includegraphics[scale=0.30]{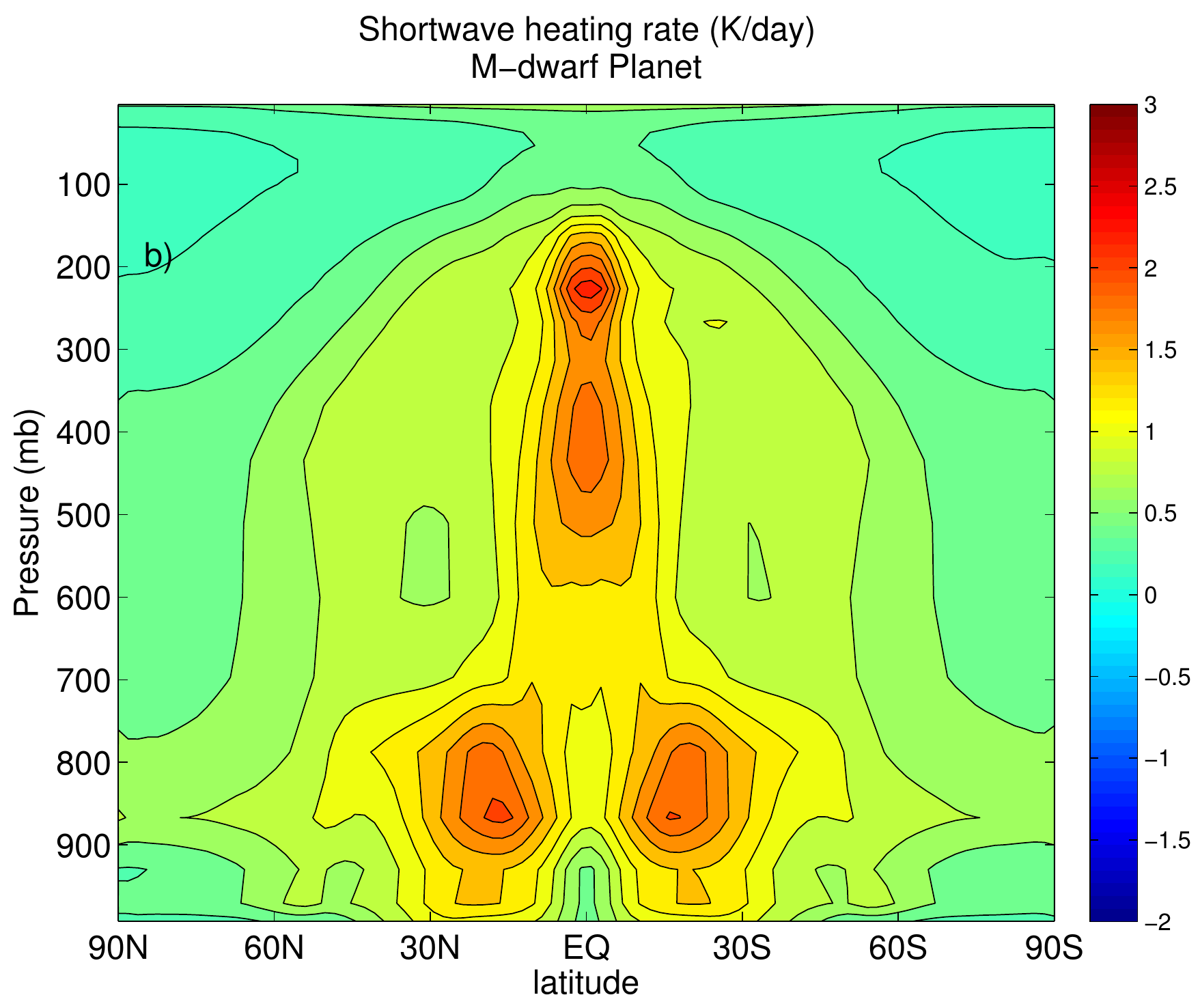}
\includegraphics[scale=0.30]{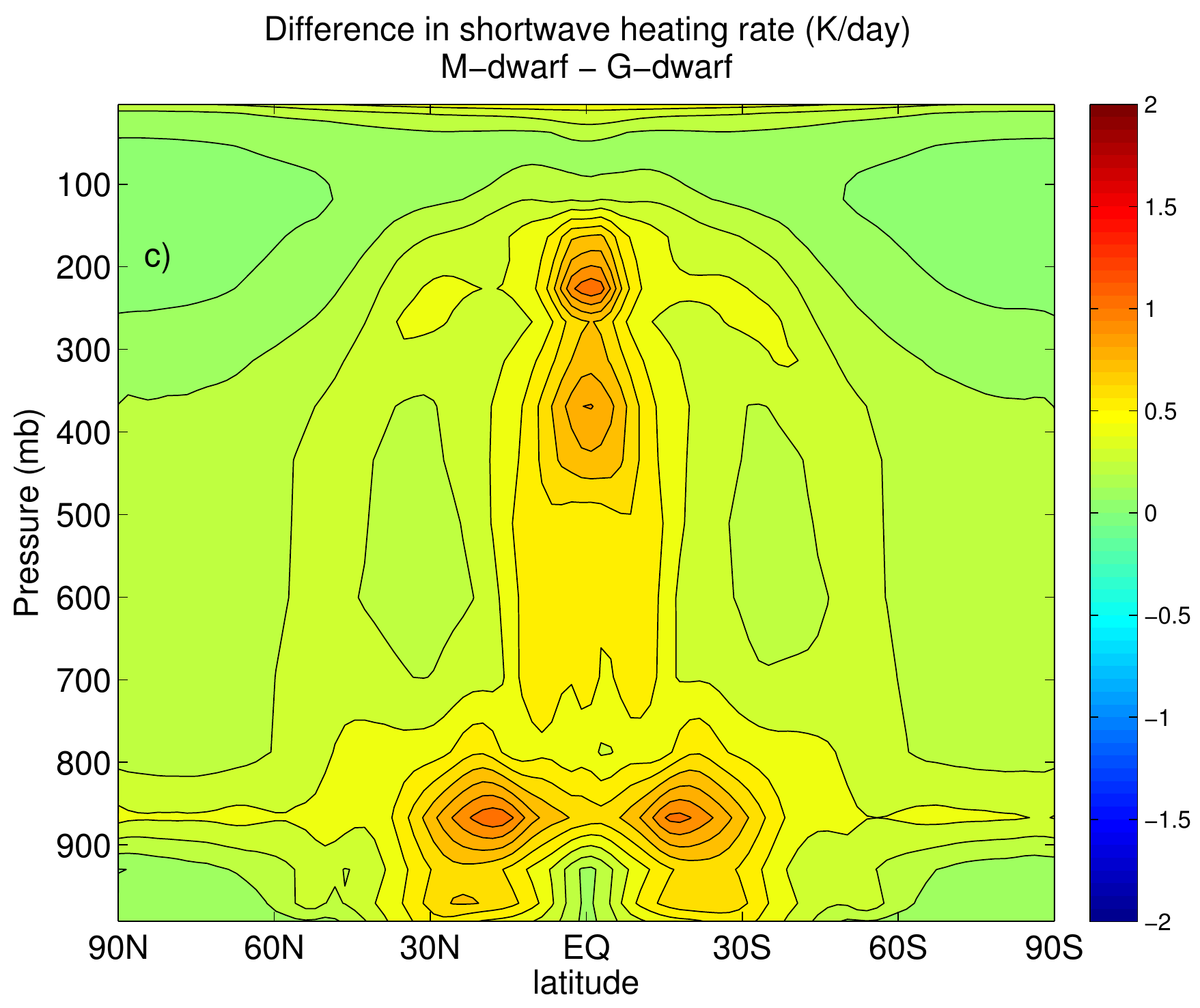}\\
\includegraphics[scale=0.30]{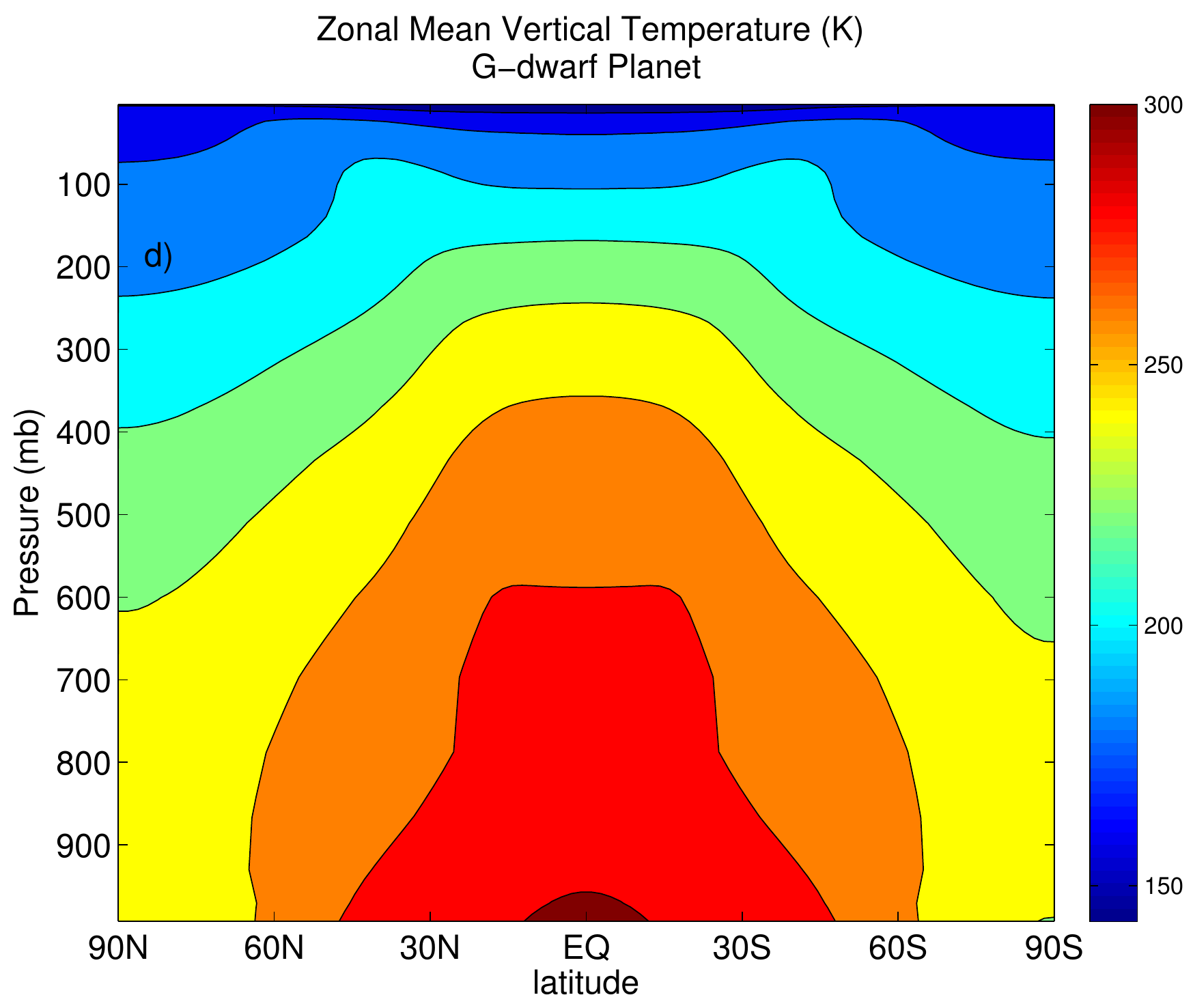}
\includegraphics[scale=0.30]{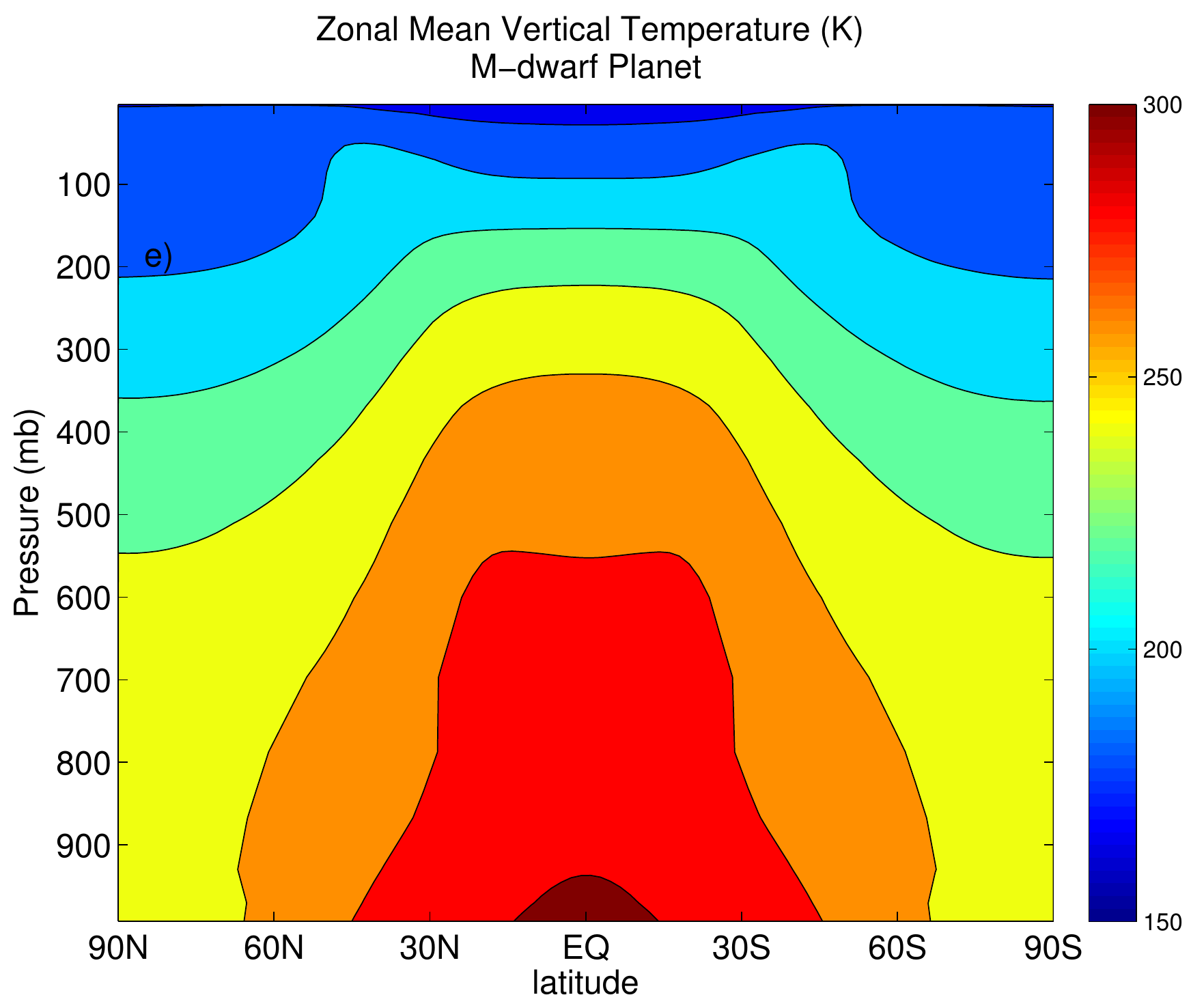}
\includegraphics[scale=0.30]{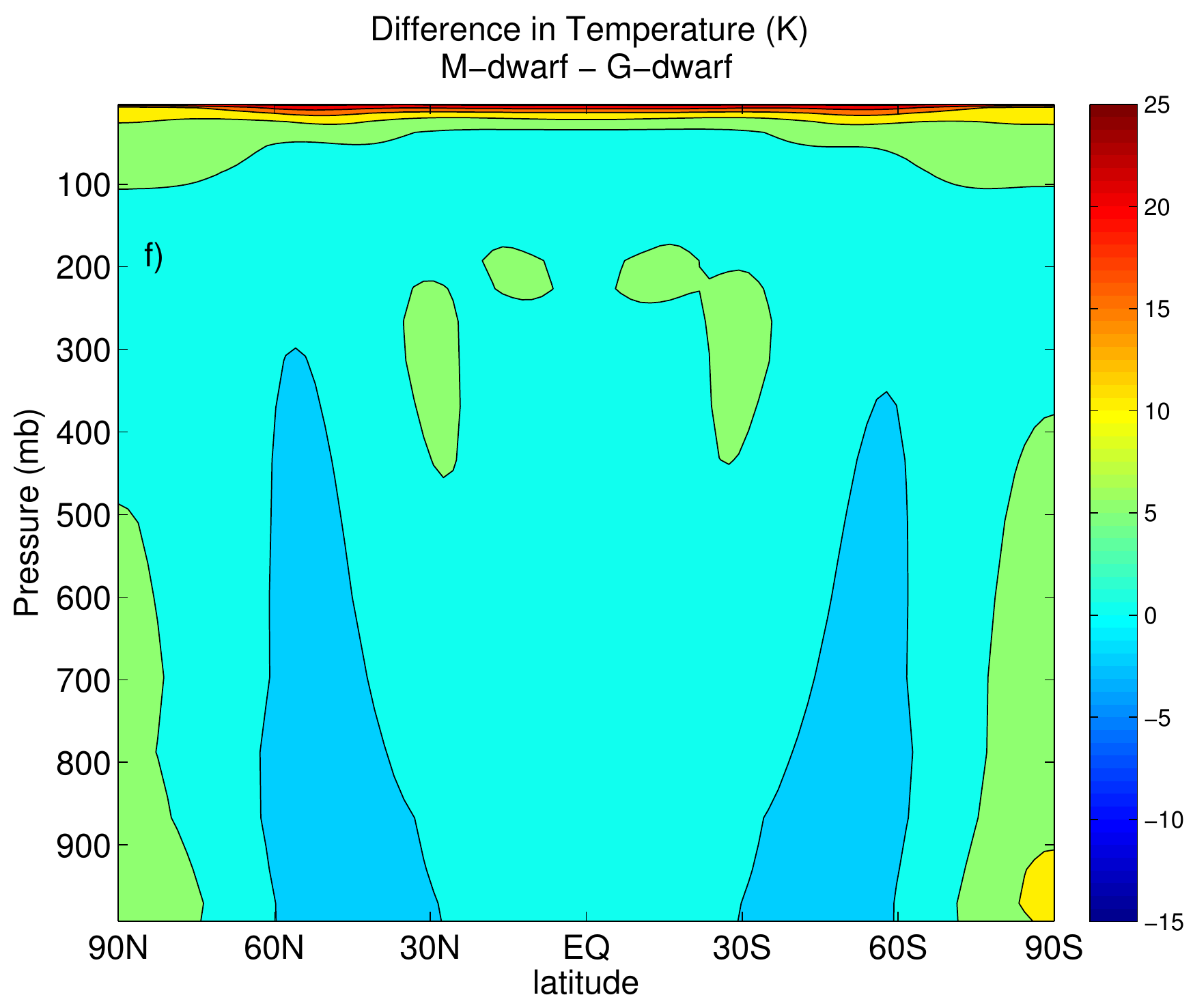}
\caption{GCM Comparison of an M-dwarf aqua planet receiving 90\% of the modern solar constant compared with an aqua planet receiving 100\% of the modern solar constant from the Sun, a G-dwarf star. (a) shortwave heating in the atmosphere of the G-dwarf planet; (b) shortwave heating in the atmosphere of the M-dwarf planet; (c) increase in shortwave heating in the atmosphere of the M-dwarf planet, calculated by taking the difference between the M-dwarf planet's shortwave heating profile and the G-dwarf planet's shortwave heating profile; (d) zonal mean temperature in the atmosphere of the G-dwarf aqua planet; (e) zonal mean temperature in the atmosphere of the M-dwarf aqua planet; (f) increase in zonal mean temperature in the atmosphere of the M-dwarf aqua planet, calculated by taking the difference between the M-dwarf planet's atmospheric temperature profile and the G-dwarf planet's atmospheric temperature profile.}
\label{Figure 11.}
\end{center}
\end{figure}
\pagebreak

\begin{figure}[!htb]
\begin{center}
\includegraphics[scale=0.40]{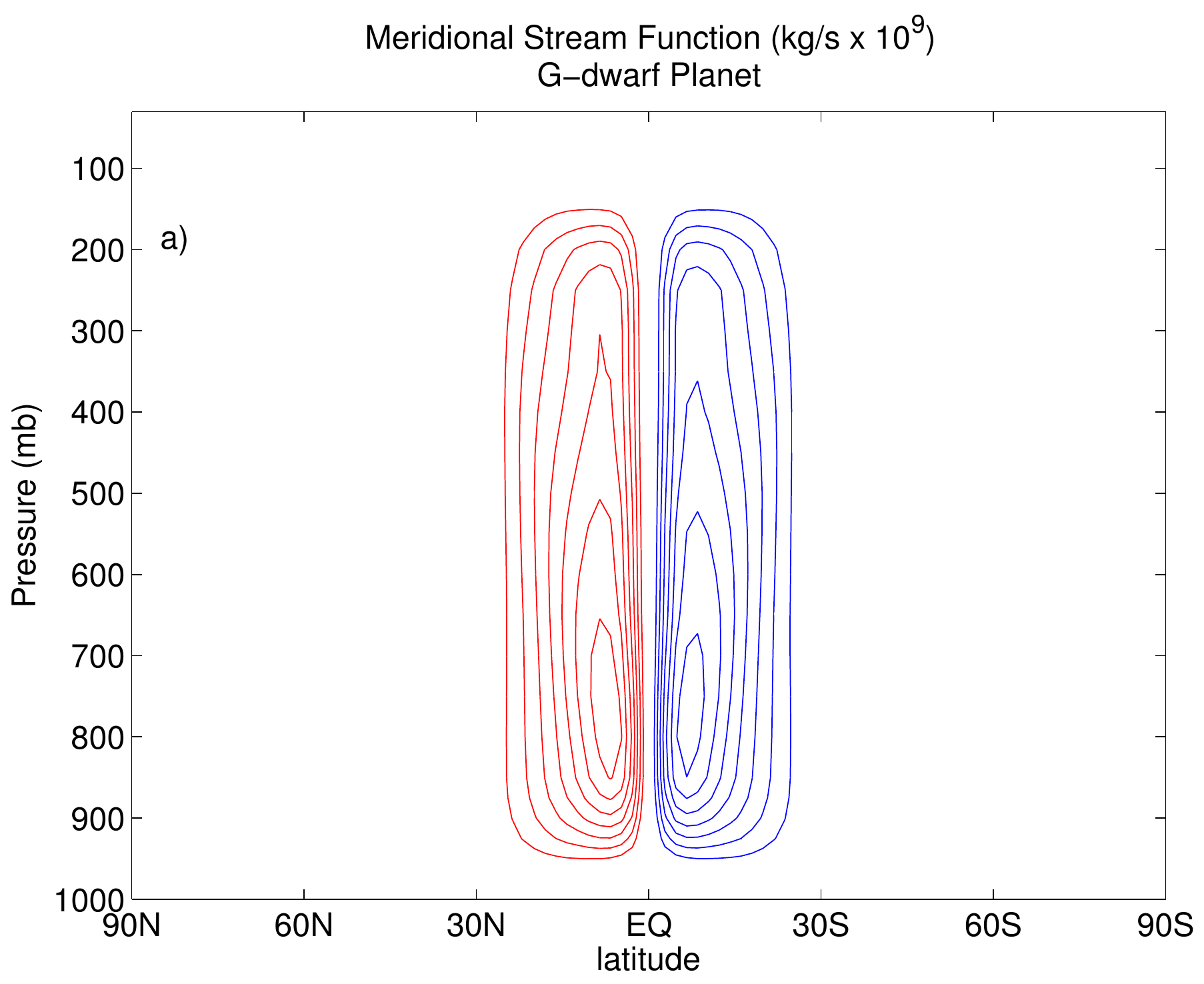}
\includegraphics[scale=0.40]{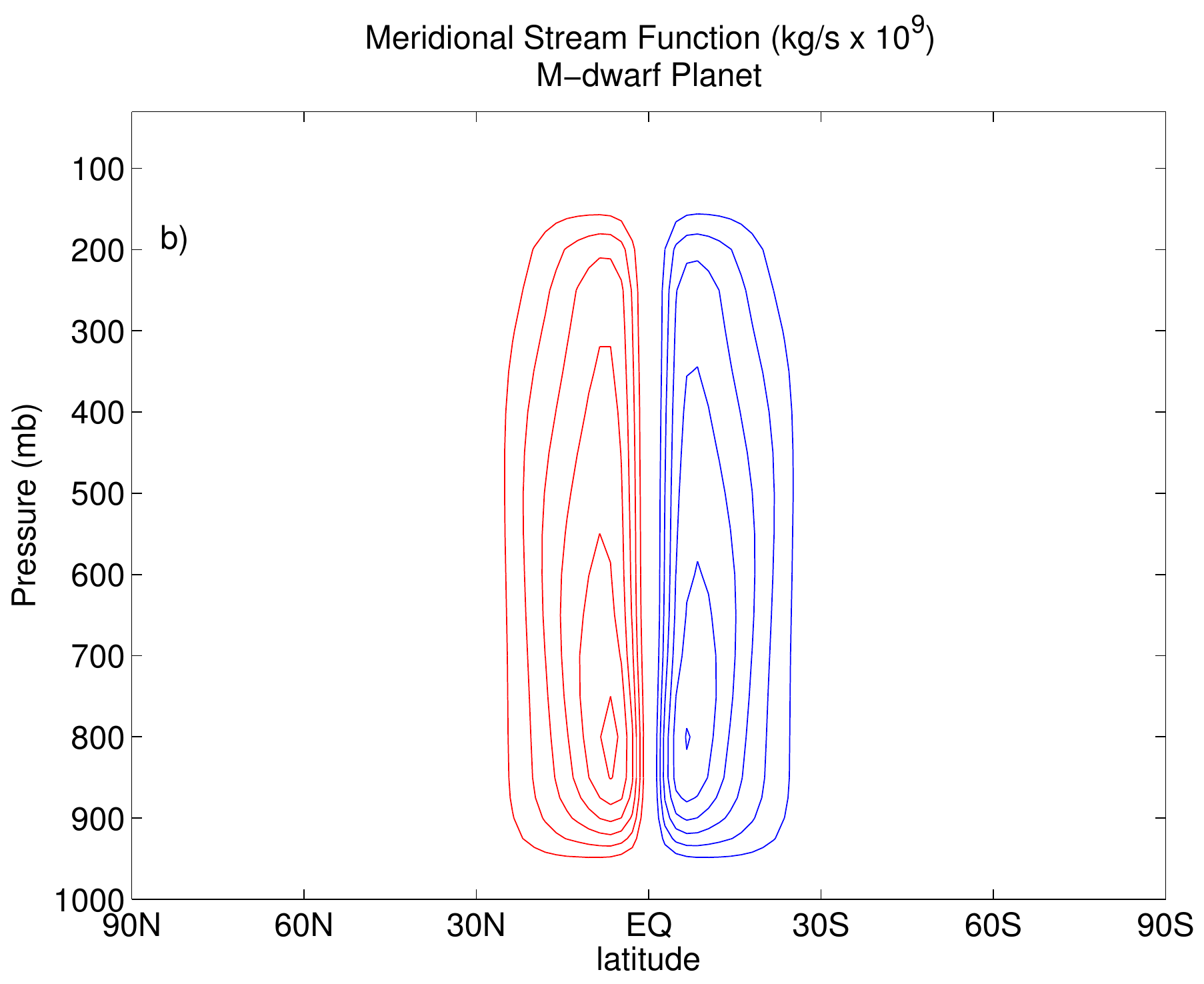}
\includegraphics[scale=0.40]{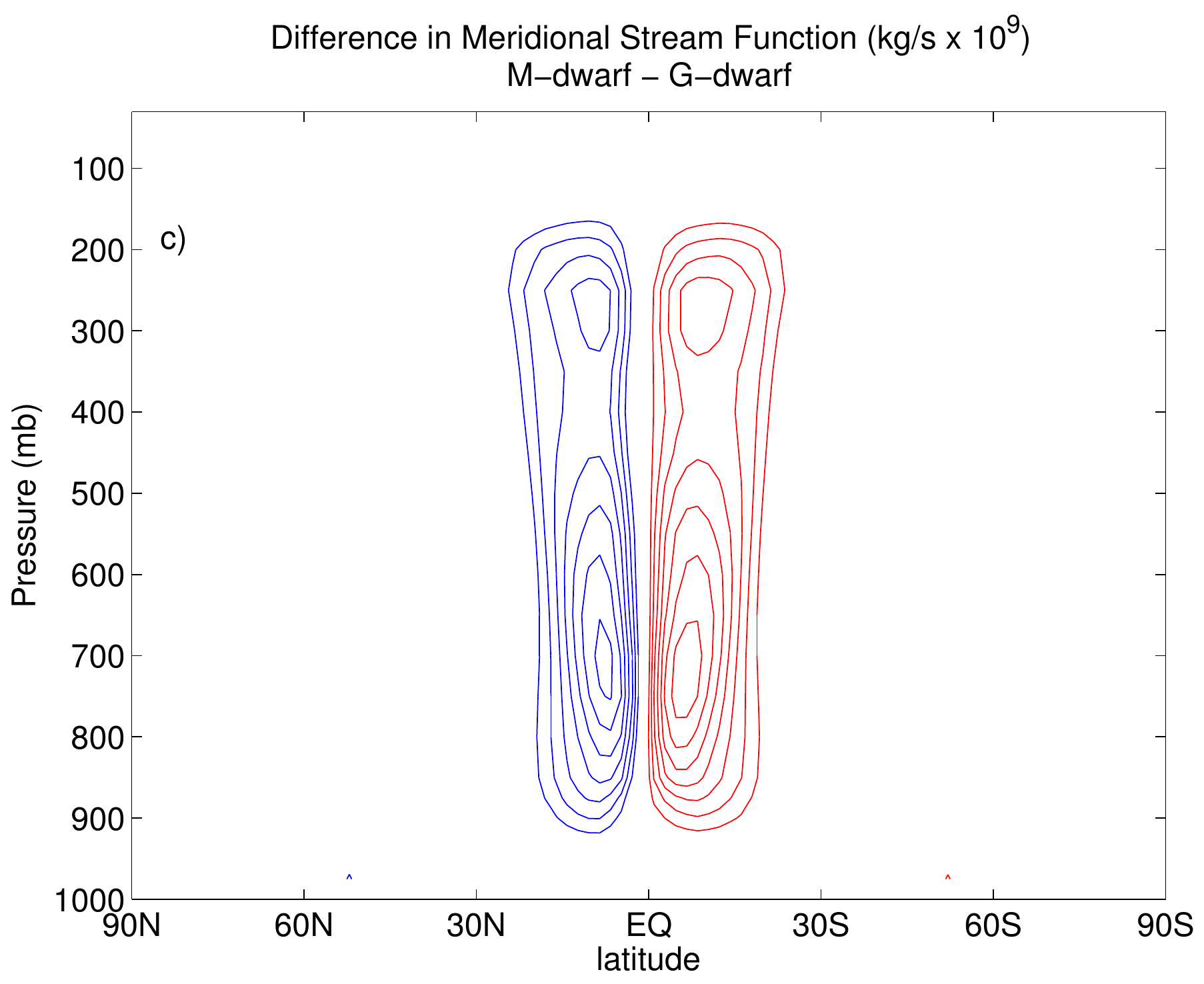}\\
\caption{GCM Comparison of an M-dwarf aqua planet receiving 90\% of the modern solar constant compared with an aqua planet receiving 100\% of the modern solar constant from the Sun, a G-dwarf star. Blue = clockwise circulation. Red = counterclockwise circulation. (a) Meridional stream function in the atmosphere of the G-dwarf planet. The contours start at 50 kg/s x 10$^9$, and the contour interval is 25 kg/s x 10$^9$; (b) Meridional stream function in the atmosphere of the M-dwarf planet. The contours start at 50 kg/s x 10$^9$, and the contour interval is 25 kg/s x 10$^9$; (c) Increase in the meridional stream function in the atmosphere of the M-dwarf planet, calculated by taking the difference between the M-dwarf planet's meridional stream function and the G-dwarf planet's meridional stream function. The contours start at 10 kg/s x 10$^9$, and the contour interval is 5 kg/s x 10$^9$. The weaker Hadley circulation on the M-dwarf planet results in greater atmospheric temperatures, and less heat transported away from the equator, compensating for the reduced instellation relative to the G-dwarf planet.}
\label{Figure 12.}
\end{center}
\end{figure}

\begin{figure}[!htb]
\begin{center}
\includegraphics[scale=0.75]{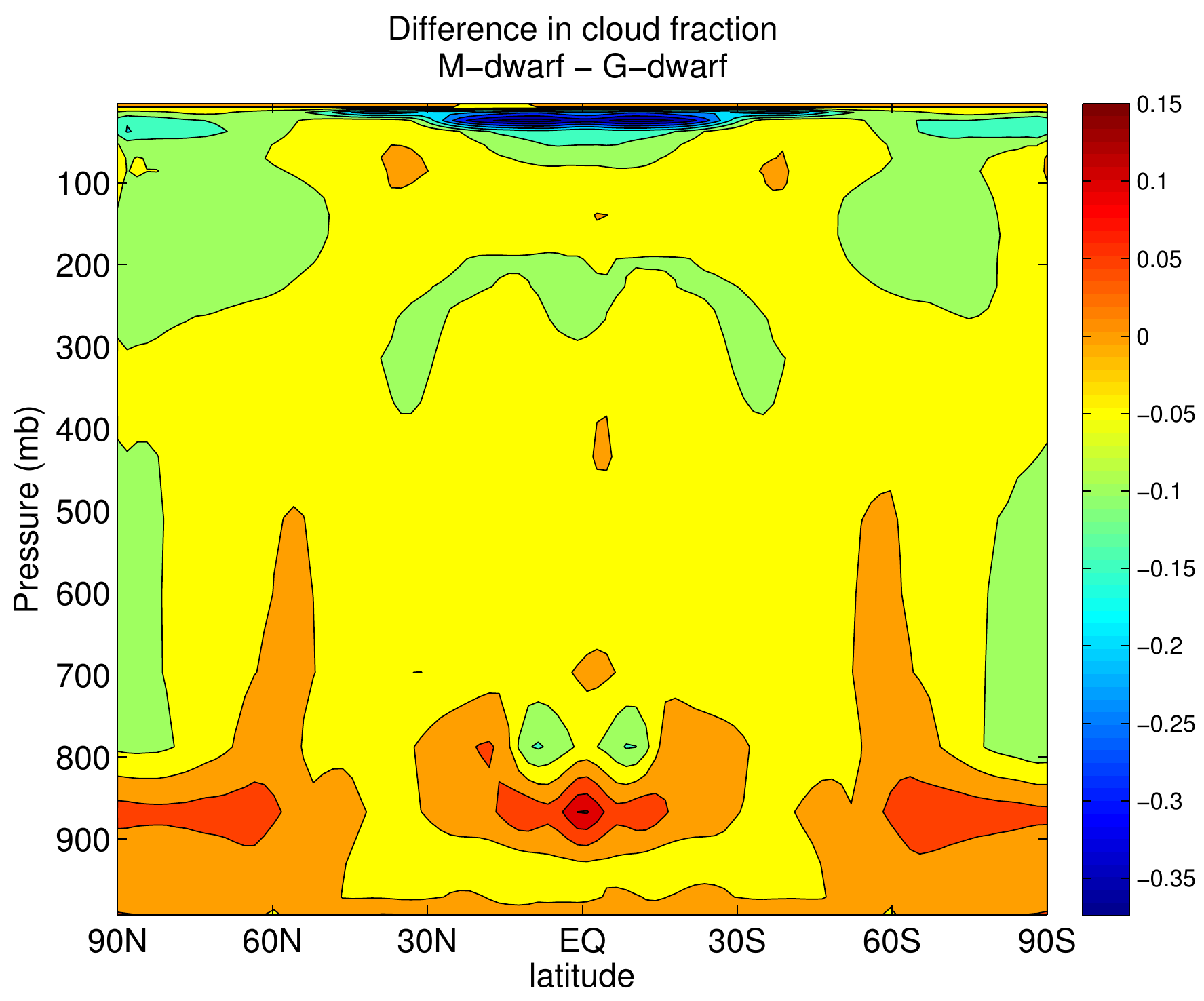}
\caption{Difference between the cloud fraction in the atmosphere of an aqua planet receiving 90\% of the modern solar constant from an M-dwarf star and an aqua planet receiving 100\% of the modern solar constant from the Sun after a 40-year GCM run.}
\label{Figure 13.}
\end{center}
\end{figure}

\begin{figure}[!htb]
\begin{center}
\includegraphics[scale=0.75]{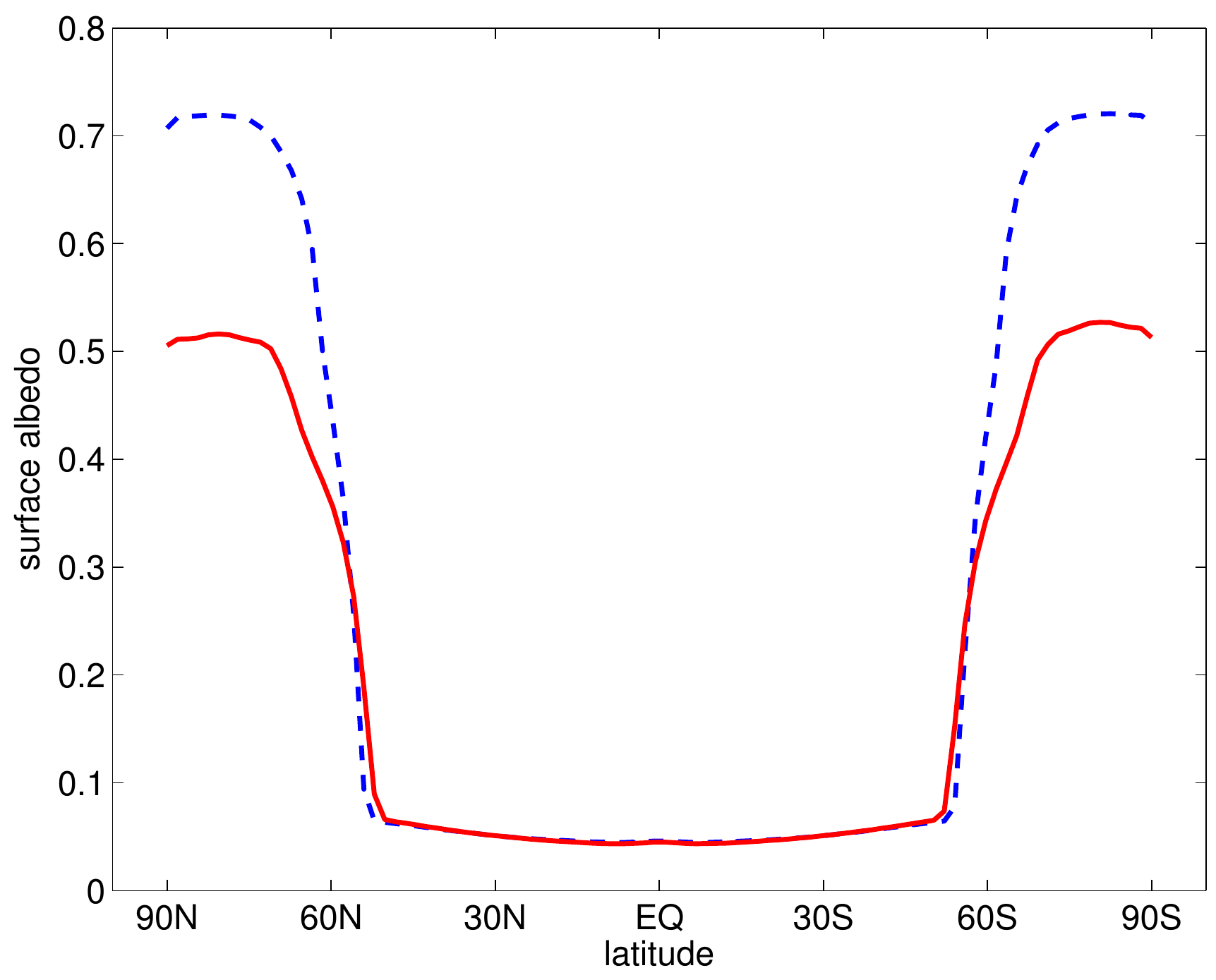}
\caption{Surface albedo as a function of latitude for an M-dwarf aqua planet receiving 90\% of the modern solar constant (red) compared with an aqua planet receiving 100\% of the modern solar constant from the Sun, a G-dwarf star (blue), after a 40-year GCM run.}
\label{Figure 14.}
\end{center}
\end{figure}

\pagebreak

\begin{figure}[!htb]
\begin{center}
\includegraphics[scale=0.40]{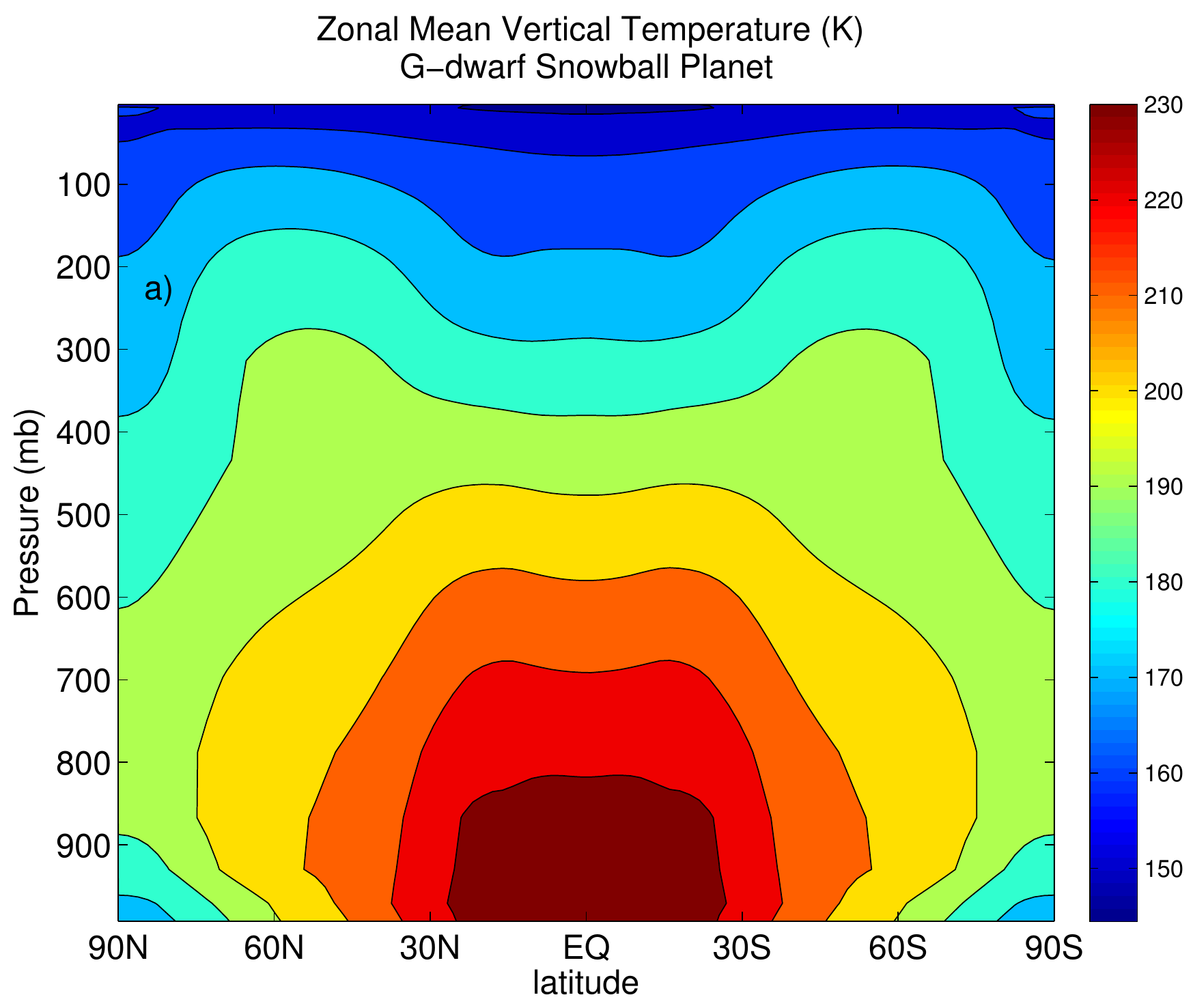}
\includegraphics[scale=0.40]{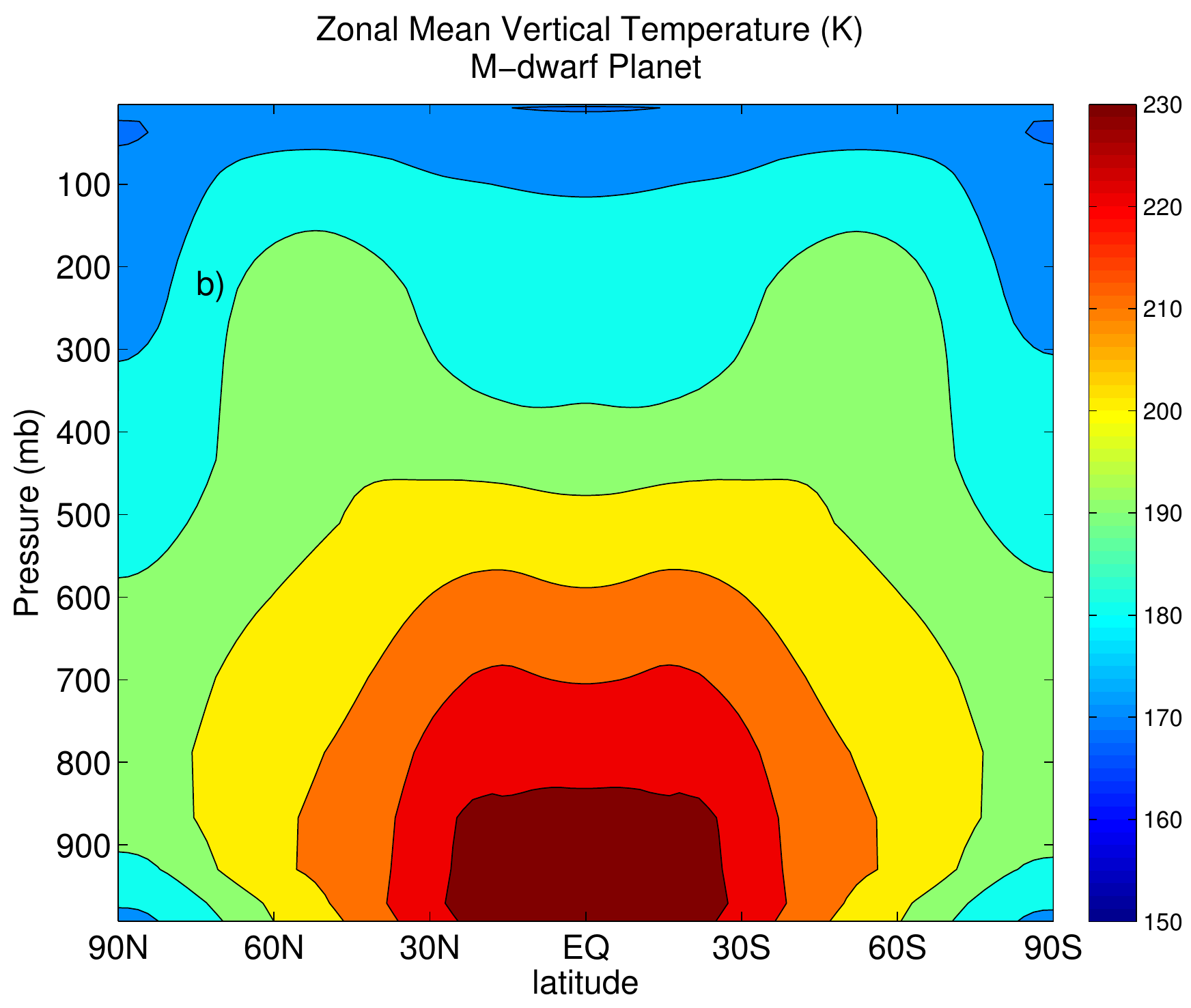}
\includegraphics[scale=0.40]{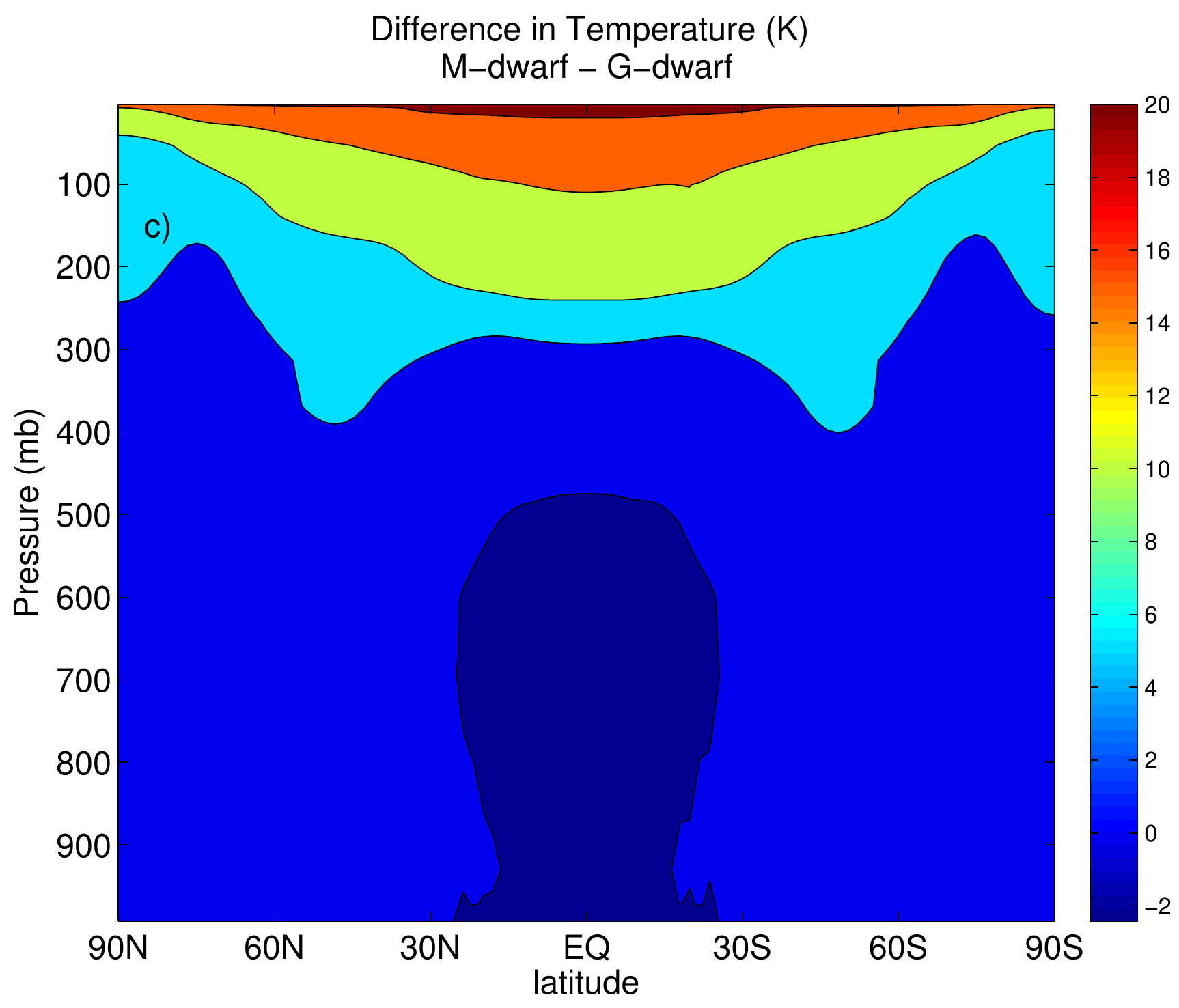}
\caption{GCM Comparison of an M-dwarf aqua planet receiving 73\% of the modern solar constant (993 W/m$^2$) with an aqua planet orbiting the Sun and receiving 92\% of the modern solar constant (1251 W/m$^2$). Both planets are completely ice-covered. (a) zonal mean temperature in the atmosphere of the G-dwarf planet; (b) zonal mean temperature in the atmosphere of the M-dwarf planet; (c) increase in temperature in the atmosphere of the M-dwarf aqua planet, calculated by taking the difference between the M-dwarf planet's atmospheric temperature profile and the G-dwarf planet's atmospheric temperature profile.}
\label{Figure 15.}
\end{center}
\end{figure}

\begin{figure}[!htb]
\begin{center}
\includegraphics[scale=0.40]{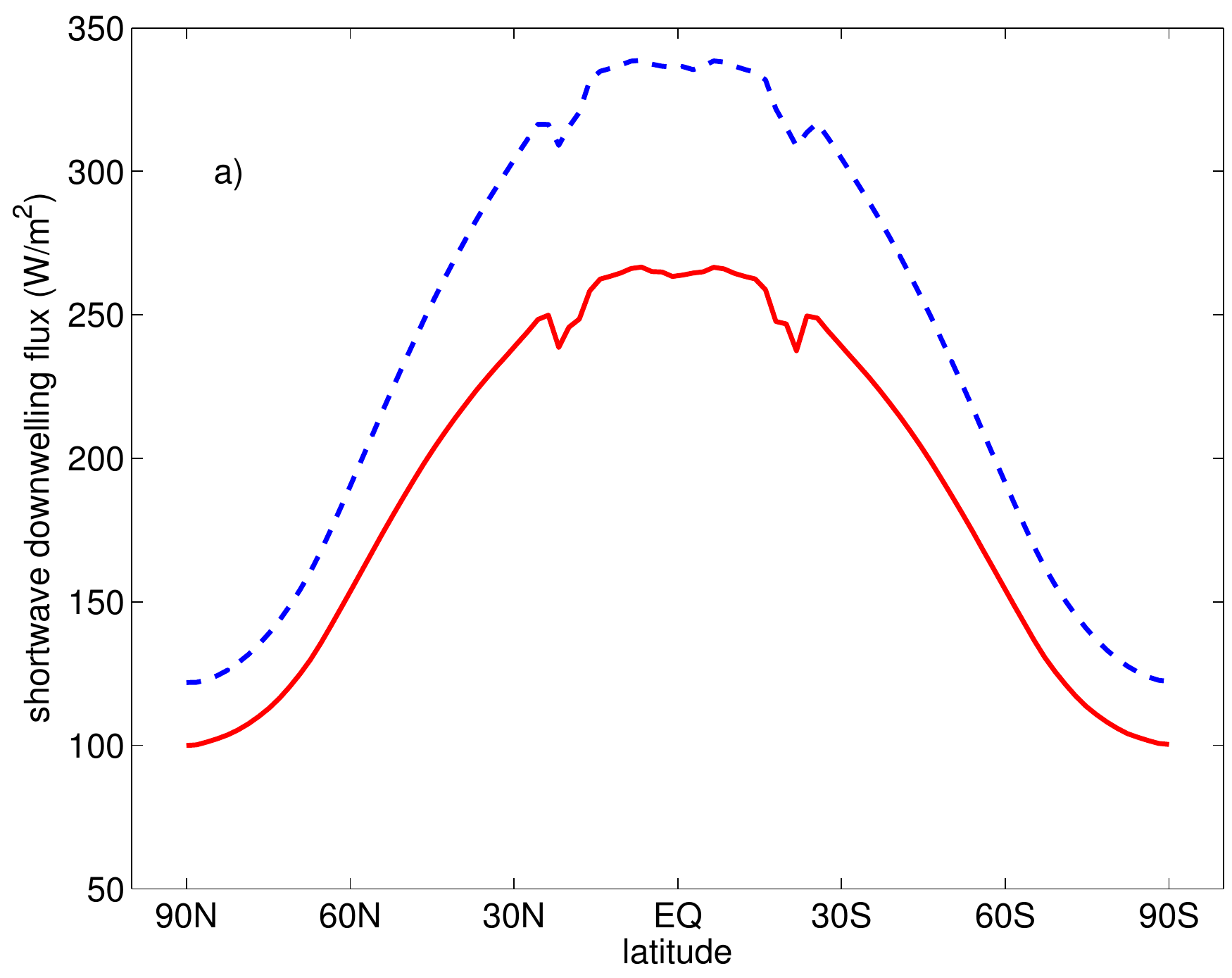}
\hspace{2 mm}
\includegraphics[scale=0.40]{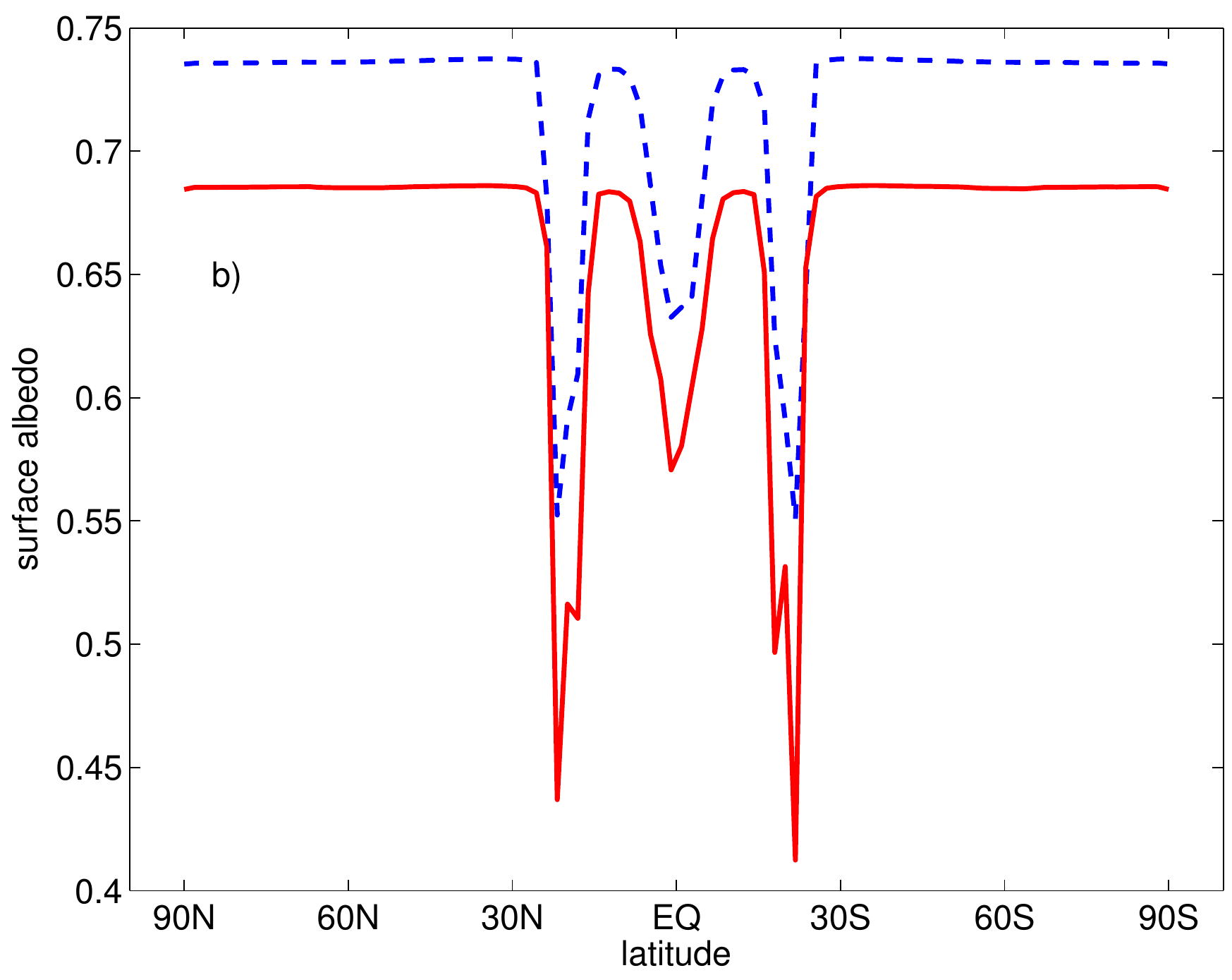}\\
\vspace{10 mm}
\includegraphics[scale=0.40]{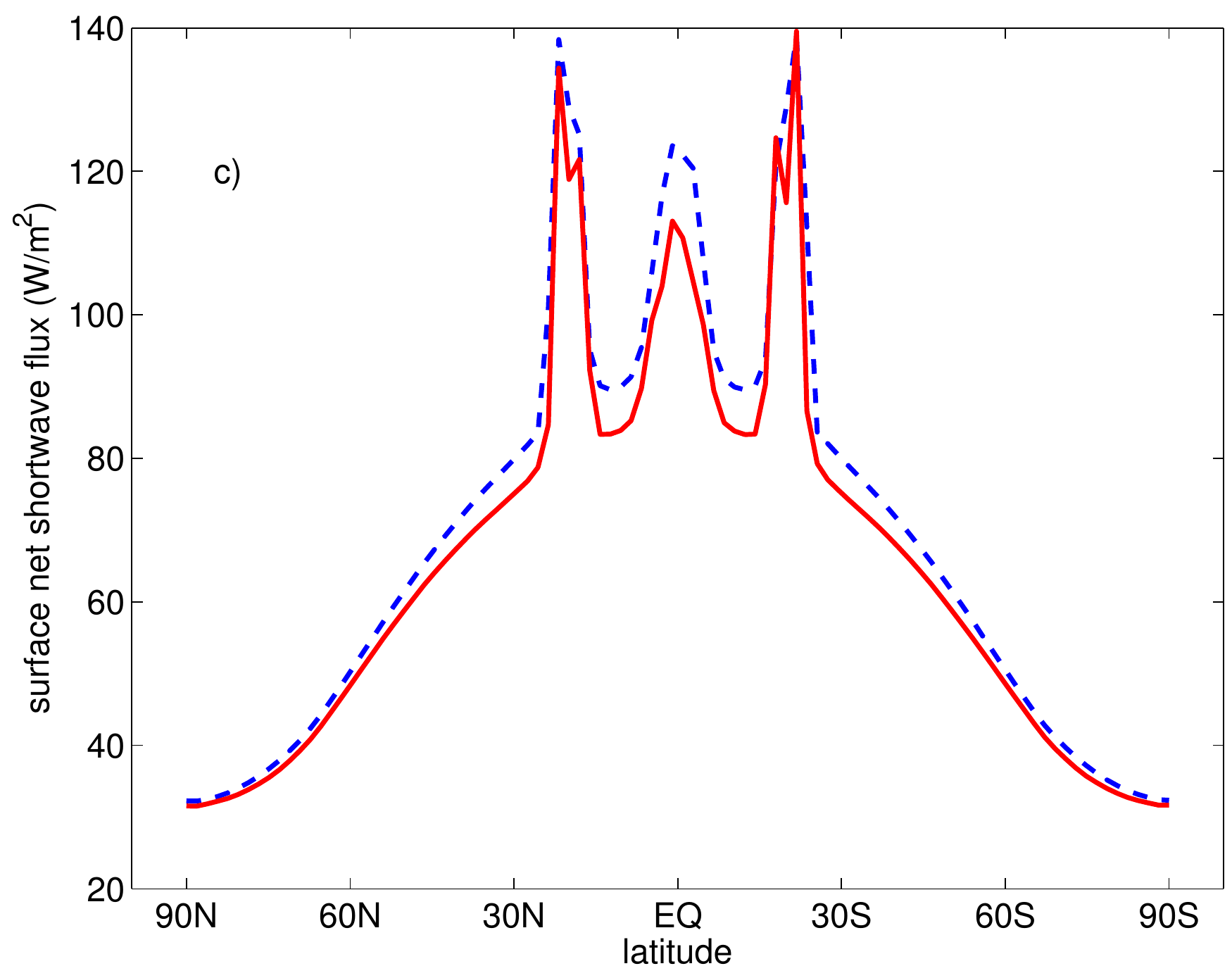}
\hspace{4 mm}
\includegraphics[scale=0.40]{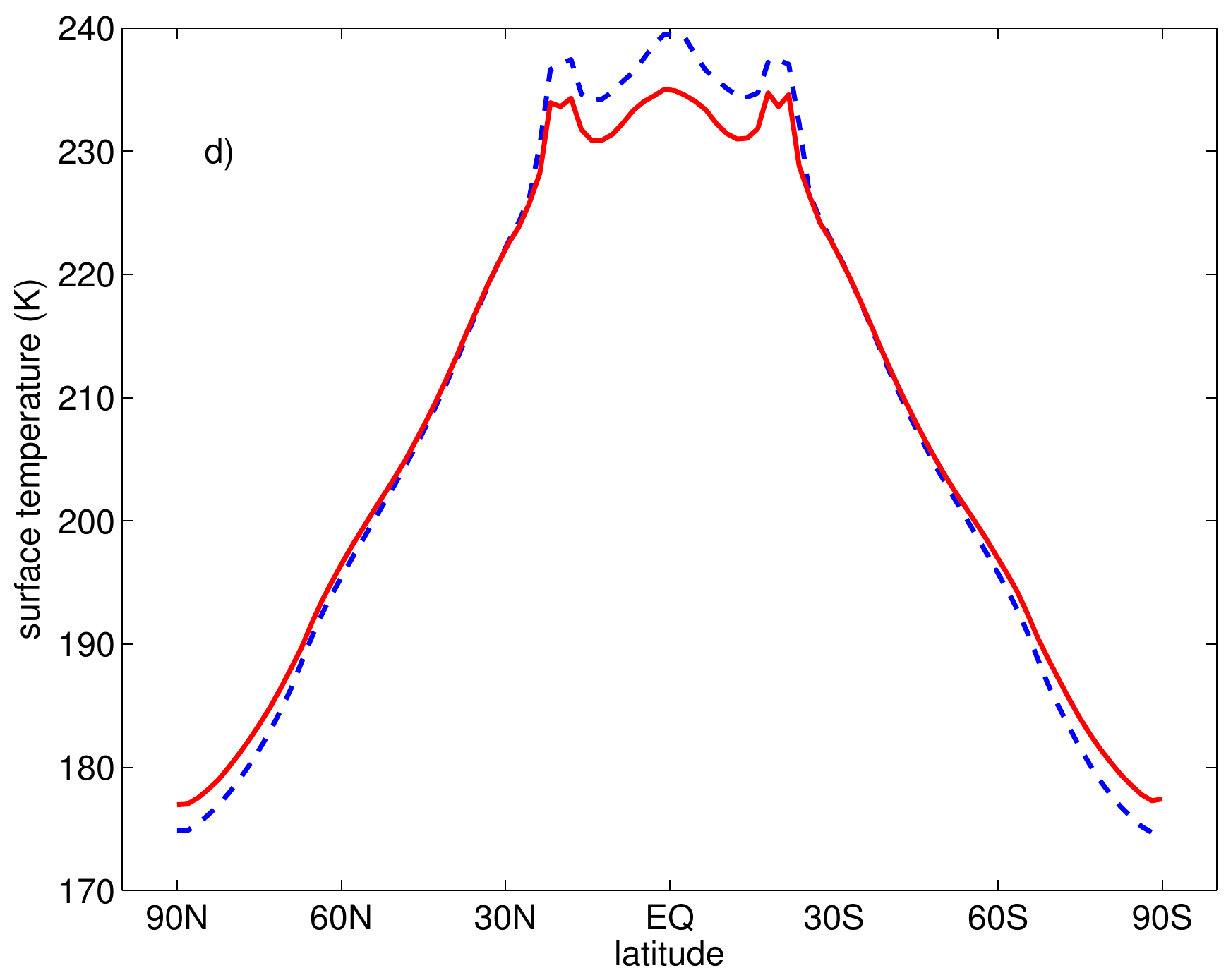}
\caption{GCM Comparison of an M-dwarf aqua planet (red) receiving 73\% of the modern solar constant (993 W/m$^2$) with an aqua planet orbiting the Sun and receiving 92\% of the modern solar constant (1251 W/m$^2$, blue). Both planets are completely ice-covered. (a) Surface shortwave downwelling flux as a function of latitude; (b) surface albedo; (c) net shortwave flux absorbed by the surface; (d) surface temperature.}
\label{Figure 16.}
\end{center}
\end{figure}

\begin{figure}[!htb]
\begin{center}
\includegraphics[scale=1.5]{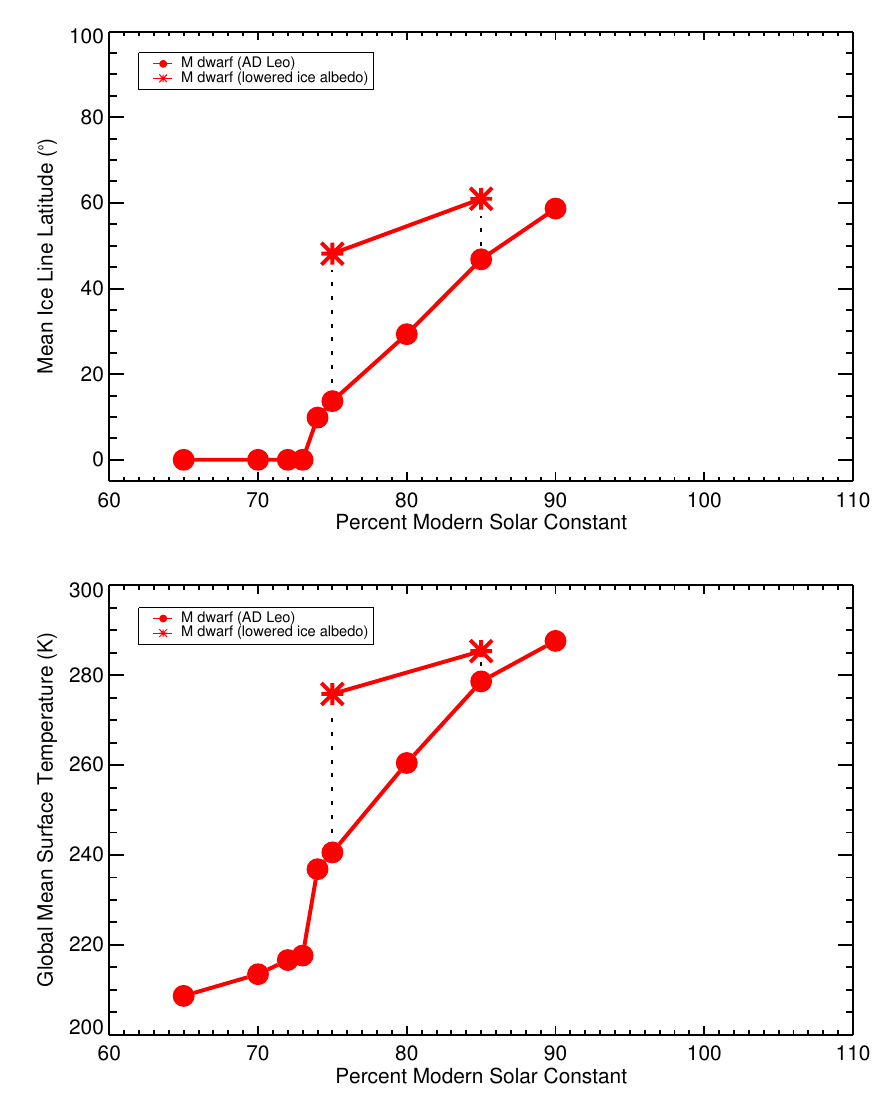}
\caption{Mean ice line latitude (top) and global mean surface temperature (bottom) as a function of percent of modern solar constant after a 40-year GCM run for an aqua planet orbiting an M-dwarf star, from Figure 8. Also plotted here are the resulting ice extents and global mean surface temperatures for M-dwarf planets receiving 75\% and 85\% instellation, with IR and visible band ice and snow albedos lowered to 0.2 (asterisks). The difference in climates is larger between the M-dwarf planets receiving 75\% instellation than between the M-dwarf planets receiving 85\% instellation (as indicated by the black vertical lines), exhibiting a shallower change in global mean surface temperature and ice extent for lowered instellation than with the default albedo parameterization.  }
\label{Figure 17.}
\end{center}
\end{figure}

\begin{figure}[!htb]
\begin{center}
\includegraphics [scale=2.00]{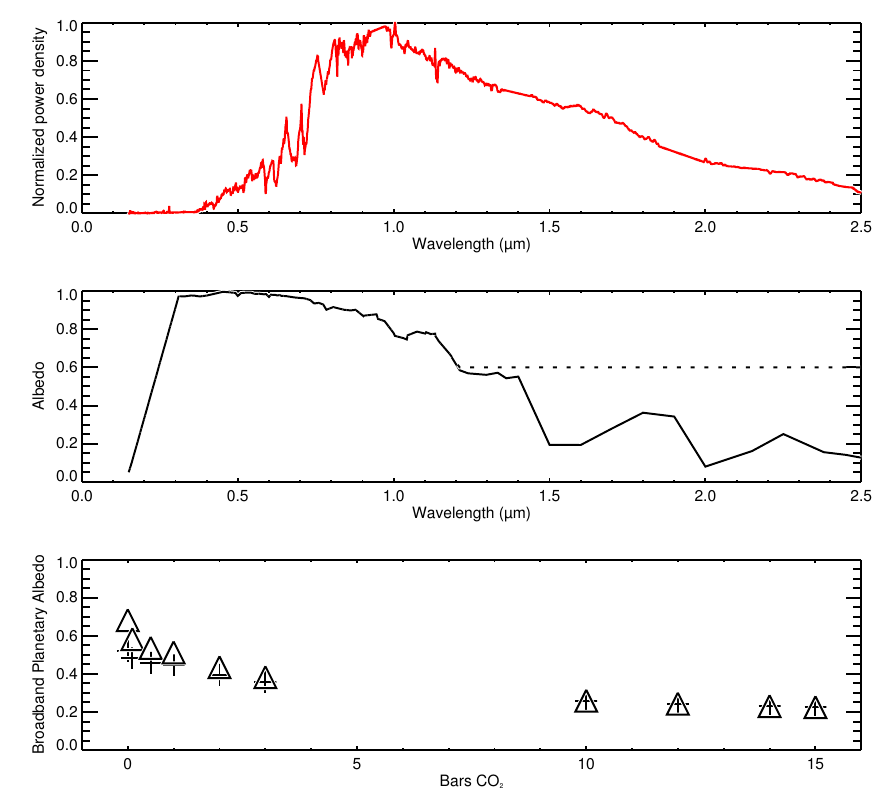}\\
\caption{Top: The normalized SED of the M3V star AD Leo. Middle: The albedo spectrum of fine-grained snow (solid line), with artificially-enhanced albedo values of 0.6 at wavelengths longer than 1.1$\mu$m (dashed line). Bottom: Broadband planetary albedos (0.15 $\mu$m $\leqslant \lambda \leqslant$ 2.5 $\mu$m) as output from SMART, given an incident M-dwarf spectrum, input actual (plus symbols) and artificially-enhanced (triangles) snow albedo spectra, and various concentrations of atmospheric CO$_2$. The concentration of CO$_2$ can be expected to effectively mask the ice-albedo spectral dependence when the broadband planetary albedo for the actual input snow spectrum (with lower near-IR albedo values) matches that for the artificially-enhanced snow spectrum (with high values of near-IR albedo), demonstrating that broadband planetary albedo is no longer sensitive to the surface albedo of the planet. This appears to happen at atmospheric concentrations of between 3 and 10 bars of CO$_2$.}
\label{Figure 18.}
\end{center}
\end{figure}

\begin{figure}[!htb]
\begin{center}
\includegraphics[scale=1.35]{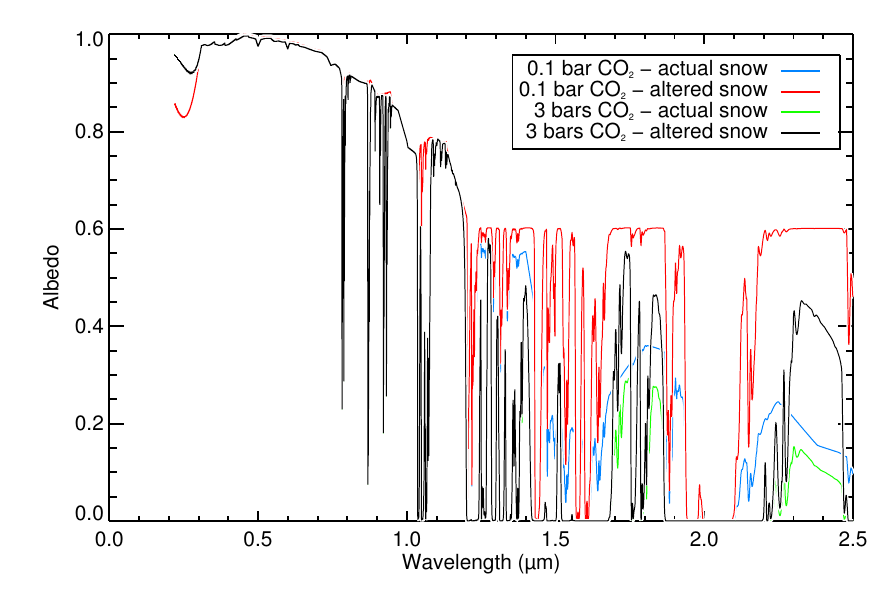}\\
\includegraphics[scale=1.35]{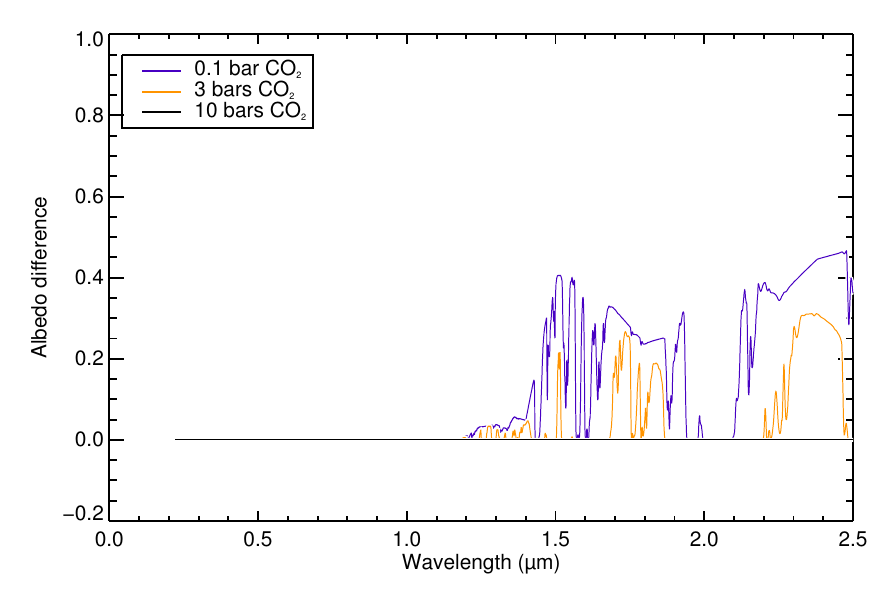}
\caption{ Wavelength-dependent reflectivity of a planet with various concentrations of atmospheric CO$_2$ and an underlying snow surface, calculated with SMART using the actual measured albedo spectrum, and one that was altered to exhibit artificially high albedo values of 0.6 at wavelengths longer than 1.1 $\mu$m. Top: 0.1 bar of CO$_2$ and an underlying snow surface matching the fine-grained snow spectrum in Figure 1 (blue); 0.1 bar of CO$_2$ with a snow surface with an artificially-enhanced spectrum (red); 3 bar-CO$_2$ atmosphere with the actual snow spectrum (green);  3 bar-CO$_2$ atmosphere with the artificially-enhanced snow spectrum (black). Bottom: Change in reflectivity between the planets with artificially-enhanced vs. actual snow surface albedo spectra. With 3 bars of CO$_2$ in the atmosphere, the difference between the albedos of the planets (orange) has decreased significantly compared to that of the 0.1-bar planets (purple). With 10 bars of CO$_2$ in the atmosphere, the difference in the albedo spectra of the planets (black) is close to zero, due to increased near-IR absorption by CO$_2$ at longer wavelengths.}
\label{Figure 19.}
\end{center}
\end{figure}

\begin{figure}[!htb]
\begin{center}
\includegraphics [scale=0.68]{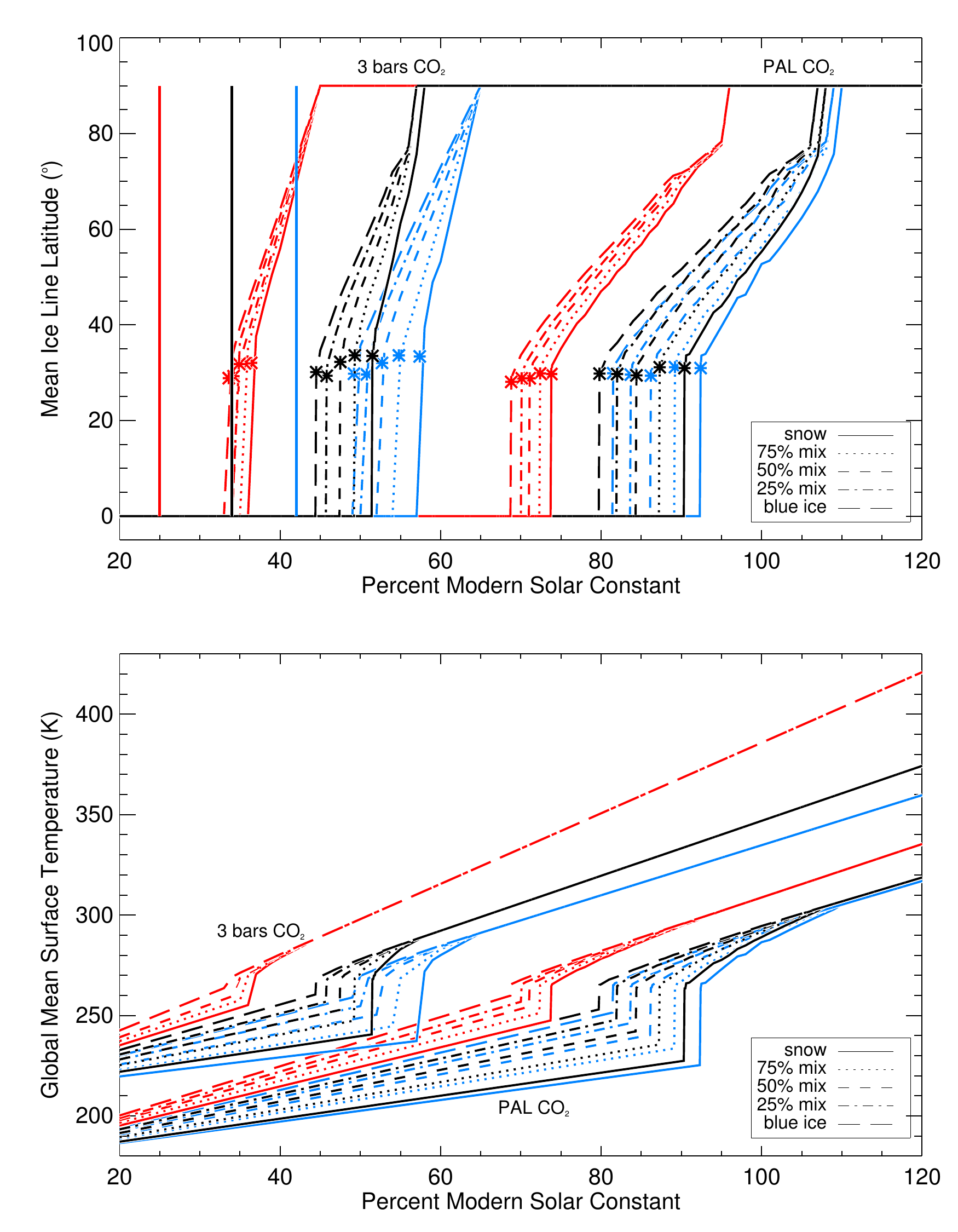}\\	
\caption{Mean ice line latitude (top) and global mean surface temperature (bottom) in the northern hemisphere as a function of percent of modern solar constant are calculated using a seasonal EBM, at present Earth obliquity (23.5$^\circ$) for aqua planets (land and ocean fraction 0.01 and 0.99, respectively) orbiting F-, G-, and M-dwarf stars at an equivalent flux distance, as in Figure 6. Here the present atmospheric level (PAL) of CO$_2$, as well as 3 bars of CO$_2$ were used (F-dwarf planet in blue, G-dwarf planet in black, and M-dwarf planet in red). Asterisks denote the minimum ice line latitude before collapse to the equator and global ice coverage. Also plotted here as vertical solid lines are the updated maximum CO$_2$ greenhouse limits for the F-dwarf (blue), G-dwarf (black), and M-dwarf (red) planets \citep{Kopparapu2013}.}
\label{Figure 20.}
\end{center}
\end{figure}

\end{document}